\documentclass{aa}

\usepackage{natbib}
\usepackage{graphicx}
\usepackage{amsmath}
\usepackage{amsbsy}
\usepackage{txfonts}

\usepackage{color}
\definecolor{Red}{rgb}{1.0,0.0,0.0}

\newcommand{\q}[1]{_{\rm #1}}
\newcommand{\ten}[1]{$10^{#1}$}
\newcommand{\scit}[2]{$#1\times10^{#2}$}

\newcommand{\ps}{s$^{-1}$}
\newcommand{\pcs}{cm$^{-2}$}
\newcommand{\pcc}{cm$^{-3}$}

\newcommand{\kkms}{K km s$^{-1}$}
\newcommand{\micron}{$\mu$m}

\newcommand{\fig}[1]{Fig.\ \ref{fig:#1}}
\newcommand{\figg}[1]{Figure \ref{fig:#1}}
\newcommand{\tb}[1]{Table \ref{tb:#1}}
\newcommand{\sect}[1]{Sect.\ \ref{sec:#1}}
\newcommand{\sectt}[1]{Section \ref{sec:#1}}
\newcommand{\nodata}{$\cdots$}
\newcommand{\mcnodata}{\multicolumn{2}{c}{\nodata}}
\newcommand{\mh}{H$_2$}
\newcommand{\mhm}{{\rm H}_2}
\newcommand{\hcop}{HCO$^+$}
\newcommand{\htcop}{H$^{13}$CO$^+$}
\newcommand{\cdo}{CO$_2$}
\newcommand{\mn}{N$_2$}
\newcommand{\nnhp}{N$_2$H$^+$}
\newcommand{\amh}{NH$_3$}
\newcommand{\meh}{CH$_4$}
\newcommand{\w}{H$_2$O}
\newcommand{\ceo}{C$^{18}$O}
\newcommand{\eup}{E_{\rm u}}
\newcommand{\tmbdv}{\int T_{\rm mb}{\rm d}v}

\defcitealias{visser12b}{Paper I}
\newcommand{\vb}{\citetalias{visser12b}}

\begin{document}

\title{Chemical tracers of episodic accretion in low-mass protostars\thanks{Herschel is an ESA space observatory with science instruments provided by European-led Principal Investigator consortia and with important participation from NASA.}}

\author{
Ruud Visser\inst{1,2}
\and Edwin A. Bergin\inst{2}
\and Jes K. J{\o}rgensen\inst{3}
}

\institute{
European Southern Observatory, Karl-Schwarzschild-Str.\ 2, D-85748, Garching, Germany; \email{rvisser@eso.org}
\and
Department of Astronomy, University of Michigan, 1085 S.\ University Ave., Ann Arbor, MI 48109-1107, USA
\and
Centre for Star and Planet Formation, Niels Bohr Institute \& Natural History Museum of Denmark, University of Copenhagen, {\O}ster Voldgade 5--7, DK-1350 Copenhagen K., Denmark
}

\date{Draft version \today}

\abstract
{} 
{Accretion rates in low-mass protostars can be highly variable in time. Each accretion burst is accompanied by a temporary increase in luminosity, heating up the circumstellar envelope and altering the chemical composition of the gas and dust. This paper aims to study such chemical effects and discusses the feasibility of using molecular spectroscopy as a tracer of episodic accretion rates and timescales.} 
{We simulate a strong accretion burst in a diverse sample of 25 spherical envelope models by increasing the luminosity to 100 times the observed value. Using a comprehensive gas-grain network, we follow the chemical evolution during the burst and for up to \ten{5} yr after the system returns to quiescence. The resulting abundance profiles are fed into a line radiative transfer code to simulate rotational spectra of \ceo, \hcop, \htcop, and \nnhp{} at a series of time steps. We compare these spectra to observations taken from the literature and to previously unpublished data of \hcop{} and \nnhp{} 6--5 from the Herschel Space Observatory.} 
{The bursts are strong enough to evaporate CO throughout the envelope, which in turn enhances the abundance of \hcop{} and reduces that of \nnhp. After the burst, it takes \ten{3}--\ten{4} yr for CO to refreeze and for \hcop{} and \nnhp{} to return to normal. The \w{} snowline expands outwards by a factor of $\sim$10 during the burst; afterwards, it contracts again on a timescale of \ten{2}--\ten{3} yr. The chemical effects of the burst remain visible in the rotational spectra for as long as \ten{5} yr after the burst has ended, highlighting the importance of considering luminosity variations when analyzing molecular line observations in protostars. The spherical models are currently not accurate enough to derive robust timescales from single-dish observations. As follow-up work, we suggest that the models be calibrated against spatially resolved observations in order to identify the best tracers to be used for statistically significant source samples.} 
{} 


\maketitle


\section{Introduction}
\label{sec:intro}
Stars gain most of their mass while deeply embedded in the molecular cloud core out of which they first formed. The accretion rate of matter from the collapsing core onto the star is probably highly variable in time. On cloud scales, turbulence can induce order-of-magnitude variations in the accretion rates onto individual cores \citep{padoan14a}. Within a given core, variations of similar magnitude are possible from mass-loading onto an embedded disk or pseudo-disk \citep{zhu09a,vorobyov10b}. The disk becomes unstable at some point, resulting in a short-lived accretion burst onto the star. Each burst causes a temporary increase in luminosity, heating up the circumstellar envelope and altering the chemical composition of the gas and dust. This paper analyzes such chemical effects and explores how to use molecular lines as a probe of variable accretion rates and timescales.

Variable or episodic accretion is the leading explanation for the wide range of protostellar luminosities observed in large-scale surveys \citep{kenyon90a,evans09a,kryukova12a,fischer13a,dunham14a} and for FUor and EXor outbursts in more evolved pre-main-sequence stars \citep{herbig77a,audard14a}. Episodic accretion can also affect processes such as disk fragmentation \citep{stamatellos11a,stamatellos12a} and lithium depletion \citep{baraffe10a}. It is therefore important to understand the magnitude--frequency distribution and determine how often bursts of a certain intensity occur. Based on large-scale variability surveys, the strongest bursts -- with accretion rates of at least 100 times the quiescent value -- happen every 5--50 kyr per protostar \citep{scholz13a}. The spacings between periodic shocks in jets and outflows support quiescent intervals anywhere from \ten{4} yr down to 10 yr, separating bursts with an unknown range of accretion rates \citep{devine97a,raga02a,arce13a}. High-cadence photometric surveys show luminosity variations at the 5--50\% level on timescales of hours to weeks \citep{billot12a,stauffer14a}, though only some of that may be related to accretion variability \citep{morales11a,rebull14a}. Despite the incomplete statistics, it appears that stronger accretion bursts occur less frequently than weaker ones. Such a distribution is supported by numerical simulations \citep{zhu09a,vorobyov10b}.

Direct observations of luminosity bursts were traditionally limited to the relatively evolved Class II sources, where the lack of an obscuring dusty envelope makes the bursts visible at optical wavelengths. However, bursts have now also been detected in at least five embedded Class I protostars \citep{audard14a} and even in one deeply embedded Class 0 source \citep{safron15a}. The detection of luminosity flares during the earliest phases of star formation lends credence to the notion of episodic accretion as a wide-spread phenomenon.

Accretion bursts have the ability to alter the chemical composition of the circumstellar material. For example, the 2008 burst in EX Lup sparked the production of crystalline silicates \citep{abraham09a} and boosted the column densities of mid-infrared \w{} and OH lines \citep{banzatti12a}. In very low-luminosity embedded protostars, whose envelopes are too cold to produce pure \cdo{} ice, the detection of double-peaked 15-\micron{} absorption profiles is attributed to a higher luminosity at some unknown point in the past (\citealt{kim11a,kim12a}; see also \citealt{poteet13a}).

The chemical effects of embedded accretion bursts primarily result from changes in the temperature of the circumstellar gas and dust. When a protostar enters a burst, the increase in luminosity can alter the D/H ratio of water and other species \citep{owen15a}. The higher temperatures also lead to the evaporation of some of the icy grain mantles (\citealt{lee07a}; \citealt{visser12b}, hereafter Paper I; \citealt{vorobyov13a}). After the burst ends, the luminosity returns to the quiescent value and the envelope cools down almost instantaneously \citep{johnstone13a}. However, at typical envelope densities, it takes \ten{3}--\ten{5} yr for the gas to freeze back onto the cold dust grains. The abundance profiles are out of equilibrium with the observed luminosity for all that time. This could explain the presence of spatially extended \ceo{} in eight out of 16 low-mass protostars \citep{jorgensen15a}.

The best example to date of episodic accretion chemistry in an embedded protostar is the central gap discovered in spatially resolved observations of \htcop{} $J=4$--3 in IRAS 15398 \citep{jorgensen13a}. This distribution is consistent with a picture where \hcop{} and its isotopologs are destroyed by \w{} in the inner envelope when the temperature exceeds 100 K\@. However, the current temperature at the edge of the \htcop{} gap is only $\sim$30~K\@. The mismatch between the dust temperature and the \htcop{} morphology implies that IRAS 15398 was hotter in the past and that the chemistry is still adjusting to the current cold environment. Based on the freeze-out timescale of \w, the simplest explanation for that hotter past is an accretion burst that happened 100--1000 yr ago \citep{jorgensen13a}.

The aim of the current paper is twofold: to explore episodic accretion chemistry in a sample of 25 spherical envelope models, and to address how certain molecular line ratios can be used as a chronometer of when the most recent accretion burst occurred in any given source. \sectt{obs} introduces the sample and summarizes the available observations, including new Herschel spectra of \hcop{} and \nnhp{} 6--5. \sectt{model} presents the physical and chemical models. \sectt{res} discusses the chemical effects of an accretion burst and \sect{tscales} shows how to use observed line ratios to derive episodic accretion timescales. Lastly, \sect{conc} presents the main conclusions.


\section{Source sample and observations}
\label{sec:obs}
\vb{} explored the chemical aspects of episodic accretion with a set of single-point models at representative densities of \ten{5}, \ten{6}, and \ten{7} \pcc. The current paper analyzes the chemistry in a sample of observationally determined spherical protostellar envelope models, each covering a range of densities and temperatures. We use the predicted abundance profiles to simulate molecular line spectra and identify observable diagnostics.

\begin{table*}[t!]
\caption{Observed integrated intensities ($W\equiv\tmbdv$ in \kkms) and half-power beam widths ($\theta$ in arcsec).\tablefootmark{a}}
\label{tb:linedata}
\centering
\begin{tabular}{lccccccccccccc}
\hline\hline
Source & \multicolumn{2}{c}{\hcop{} 6--5} & \multicolumn{2}{c}{\htcop{} 1--0} & \multicolumn{2}{c}{\htcop{} 3--2} & \multicolumn{2}{c}{\htcop{} 4--3} & \multicolumn{2}{c}{\nnhp{} 1--0\tablefootmark{b}} & \multicolumn{2}{c}{\nnhp{} 6--5\tablefootmark{b}} & References\rule{0pt}{0.9em} \\
 & $W$ & $\theta$ & $W$ & $\theta$ & $W$ & $\theta$ & $W$ & $\theta$ & $W$ & $\theta$ & $W$ & $\theta$ & \\
\hline
L1448 MM & 1.37 & 43 & 2.0 & 43 & 1.9 & 19 & 0.92 & 21 & 12.48 & 27 & 0.19 & 41 & 1,2,3,4\rule{0pt}{0.9em} \\
NGC1333 IRAS2A & 1.36 & 43 & 1.8 & 43 & 2.1 & 19 & 2.7 & 14 & 14.2 & 40 & 0.41 & 41 & 1,2,5 \\
NGC1333 IRAS4A & 2.29 & 43 & 2.3 & 43 & 1.4 & 19 & 0.86 & 21 & 15.9 & 40 & 0.78 & 41 & 1,2,5,6 \\
NGC1333 IRAS4B & 1.88 & 43 & 2.1 & 43 & 0.57 & 19 & \mcnodata & 13.5 & 40 & 0.64 & 41 & 1,2,5 \\
L1527 & 1.06 & 43 & 2.2 & 43 & 1.1 & 19 & 0.47 & 14 & 3.91 & 27 & $<$0.030 & 41 & 1,2,4 \\
Ced110 IRS4 & \mcnodata & \mcnodata & \mcnodata & 0.38 & 18 & 4.8 & 54 & \mcnodata & 7,8 \\
BHR71 & 0.95 & 43 & \mcnodata & \mcnodata & \mcnodata & 8.1 & 54 & \mcnodata & 1,8 \\
IRAS 15398 & 0.75 & 43 & \mcnodata & 0.52 & 28 & \mcnodata & 9.4 & 54 & \mcnodata & 8,9 \\
L483 MM & 1.25 & 43 & 1.5 & 43 & 1.7 & 19 & 1.2 & 14 & 14.05 & 27 & 0.29 & 41 & 1,2,4 \\
Ser SMM1 & 4.30 & 43 & 1.56 & $c$ & 5.8 & 19 & 3.1 & 21 & 28 & 27 & 1.48 & 41 & 1,5,8,10 \\
Ser SMM4 & 2.08 & 43 & 1.13 & $c$ & 2.9 & 19 & 1.2 & 21 & 49 & 27 & \mcnodata & 1,8,10 \\
Ser SMM3 & 1.74 & 43 & $<$0.1 & $c$ & 1.6 & 19 & 0.2 & 21 & 23 & 27 & \mcnodata & 1,8,10 \\
L723 MM & 0.77 & 43 & 0.61 & 43 & 0.94 & 19 & 0.70 & 14 & 5.70 & 27 & \mcnodata & 1,2,4 \\
B335 & 0.63 & 43 & 0.24 & 24 & 0.46 & 28 & \mcnodata & 6.05 & 27 & \mcnodata & 1,4,11 \\
L1157 & 0.54 & 43 & 0.89 & 29 & 0.61 & 19 & 0.52 & 14 & 8.38 & 27 & 0.07 & 41 & 1,2,4 \\
\hline
L1489 & 0.40 & 43 & 0.96 & 43 & 0.82 & 19 & 0.61 & 14 & 0.26 & 40 & \mcnodata & 2,5\rule{0pt}{0.9em} \\
L1551 IRS5 & \mcnodata & 1.55 & 19 & 2.4 & 19 & 2.2 & 14 & 15.7 & 40 & \mcnodata & 2,12 \\
TMR1 & \mcnodata & 1.1 & 43 & 0.51 & 19 & 0.20 & 14 & 0.37 & 40 & \mcnodata & 2 \\
TMC1A & \mcnodata & 0.44 & 19 & $<$0.27 & 19 & $<$0.13 & 14 & 7.1 & 40 & \mcnodata & 2,12 \\
TMC1 & \mcnodata & 1.26 & 19 & $<$0.27 & 19 & $<$0.12 & 14 & 5.8 & 40 & \mcnodata & 2,12 \\
HH46 IRS & \mcnodata & \mcnodata & \mcnodata & 1.1 & 18 & \mcnodata & \mcnodata & 7 \\
Elias 29 & 0.28 & 43 & 0.36 & 72 & 0.09 & 28 & \mcnodata & \mcnodata & \mcnodata & 13 \\
RNO91 & \mcnodata & 0.93 & 72 & \mcnodata & \mcnodata & 11 & 27 & \mcnodata & 8,14 \\
\hline
\end{tabular}
\tablefoot{References: (1) this work; (2) \citealt{jorgensen04c}; (3) \citealt{gregersen97a}; (4) \citealt{emprechtinger09a}; (5) \citealt{sanjose14a} and \citealt{benz15a}; (6) \citealt{blake95a}; (7) \citealt{vankempen09e}; (8) \citealt{mardones97a}; (9) \citealt{gregersen00a}; (10) \citealt{hogerheijde99a}; (11) \citealt{evans05a}; (12) \citealt{onishi02a}; (13) \citealt{boogert02b}; (14) \citealt{butner95a}.
\tablefoottext{a}{Throughout the paper, Class 0 sources appear above the horizontal line and Class I sources below it. $T\q{mb}$ is the main-beam temperature, corrected for beam efficiencies as detailed in the references. Typical uncertainties on the integrated intensities are 10--25\%. Upper limits are at the $3\sigma$ level. See \citet{yildiz13a} for a compilation of \ceo{} intensities and beam sizes.}
\tablefoottext{b}{The \nnhp{} intensities are summed over all hyperfine components (see text).}
\tablefoottext{c}{Integrated intensity over a $20''\times20''$ region \citep{hogerheijde99a}.}
}
\end{table*}

The envelope models are based on the 29 low-mass embedded protostars targeted in the key program ``Water in star-forming regions with Herschel'' \citep[WISH;][]{vandishoeck11a}. Source coordinates, distances, and other basic properties are listed in Table 1 of \citet{kristensen12a}. In order to identify the best observable diagnostics, we searched the literature for observations of molecular lines and ice column densities. The most widely available tracers are \ceo{} $J=2$--1, 3--2, and 5--4 (20, 26, and 15 sources); \htcop{} 1--0, 3--2, and 4--3 (18, 17, and 16); and \nnhp{} 1--0 (21). In addition, column densities of \cdo{}, CO, and \w{} ice are available for 14, 10, and 12 sources. Lastly, we present previously unpublished spectra of \hcop{} 6--5 and \nnhp{} 6--5 (16 and 7 sources) obtained with the Heterodyne Instrument for the Far-Infrared \citep[HIFI;][]{degraauw10a} on the Herschel Space Observatory \citep{pilbratt10a}. The full data set is summarized in Tables \ref{tb:linedata} and \ref{tb:icedata}. Appendix \ref{sec:obsdetails} offers more details on the observations and the data reduction.

For lack of molecular line observations, we omit NGC1333 IRAS3, IRAS 12496, R CrA IRS5, and HH100 IRS from the WISH target list. Our final sample therefore contains 25 protostars, including IRAS 15398 \citep[\sect{intro};][]{jorgensen13a}. These sources have bolometric luminosities from 0.8 to 35.7 $L_\odot$ and span the full evolutionary sequence from deeply embedded Class 0 to late Class I \citep{lada99a}.


\section{Model description}
\label{sec:model}


\subsection{Physical framework}
\label{sec:envmod}
Spherical density and temperature profiles are available from \citet{kristensen12a} for our entire sample of 25 protostars. The density follows a power law, $n(\mhm) \propto r^{-p}$, where the slope $p$ is one of three free parameters in a grid of models from the one-dimensional continuum radiative transfer program DUSTY \citep{ivezic99a,jorgensen02a}. The other two are the mass and size of the envelope. \citeauthor{kristensen12a} compared the model grid output to the observed spectral energy distributions (SEDs) and submillimeter brightness profiles to find the best-fit parameters for each source (see their Table C.1).

The WISH target list is biased towards bright protostars, which are likely to be in a relatively active phase of accretion. Nonetheless, for the purpose of exploring the chemical effects of episodic accretion, we treat all sources as if they are currently in a quiescent phase. Alternative assumptions of some or all sources currently experiencing a high accretion rate would produce a different set of chemical models, but would not alter the conclusions in \sect{res}.

Within the best-fit models from \citet{kristensen12a}, we simulate an accretion burst by changing the stellar luminosity and rerunning DUSTY to obtain a new temperature profile. All other parameters remain constant. Our goal is to study the chemical effects of a strong burst, such as might happen at most a few times during the embedded phase. Specifically, the burst luminosity for each source is set to be 100 times higher than the current luminosity; this is the burst intensity inferred for IRAS 15398 \citep{jorgensen13a}. The simulated bursts last for 100 yr \citep{vorobyov05b}, although shorter durations down to at least 1 yr do not affect our results. After each burst ends, we follow the quiescent chemistry for a maximum of \ten{5} yr to cover the full range of potential timescales from e.g.\ \citet{scholz13a}. The models are static and ignore any dynamical effects of the envelope collapsing and dissipating on typical timescales of a few \ten{5} yr \citep{evans09a,visser11a}.

\begin{table}[t!]
\caption{Observed ice column densities ($N(X)$ in \ten{17} \pcs).}
\label{tb:icedata}
\centering
\begin{tabular}{lcccc}
\hline\hline
Source & \cdo & CO & \w & References\rule{0pt}{0.9em}  \\
\hline
L1527 & 10 & 18 & 47 & 1\rule{0pt}{0.9em} \\
Ced110 IRS4 & 12.26 & 6.87 & \nodata & 2 \\
IRAS 15398 & 52.16 & 8.35 & 148.0 & 2 \\
L723 MM & 49.0 & \nodata & \nodata & 3 \\
\hline
L1489 & 16.20 & 7.13 & 47.0 & 2\rule{0pt}{0.9em} \\
L1551 IRS5 & 11.86\tablefootmark{a} & 4.2\tablefootmark{b} & 109.0 & 4,5 \\
TMR1 & 4.92\tablefootmark{a} & \nodata & 73.8 & 4 \\
TMC1A & 10.98\tablefootmark{a} & \nodata & 53.1 & 4 \\
TMC1 & 12.71\tablefootmark{a} & \nodata & 79.2 & 4 \\
HH46 IRS & 21.58 & 7.98 & 77.9 & 2 \\
GSS30 IRS1 & 3.28 & 2.53 & 15.3 & 2 \\
Elias 29 & 6.7 & 1.7 & 30.4 & 6,7 \\
Oph IRS63 & 6.84 & 3.49 & 20.4 & 2 \\
RNO91 & 11.66 & 7.46 & 39.0 & 2 \\
\hline
\end{tabular}
\tablefoot{References: (1) \citealt{aikawa12b}; (2) \citealt{pontoppidan08a}; (3) \citealt{cook11a}; (4) \citealt{zasowski09a}; (5) \citealt{tielens91a}; (6) \citealt{boogert00a}; (7) \citealt{boogert08a}.
\tablefoottext{a}{These values are scaled up by a factor of 1.36 relative to the published values from \citet{zasowski09a}, as recommended by \citet{cook11a} to account for different \cdo{} laboratory band strengths.}
\tablefoottext{b}{Scaled up by a factor of 1.4 to account for different band strengths.}
}
\end{table}

\begin{figure}[t!]
\centering
\includegraphics[width=\hsize]{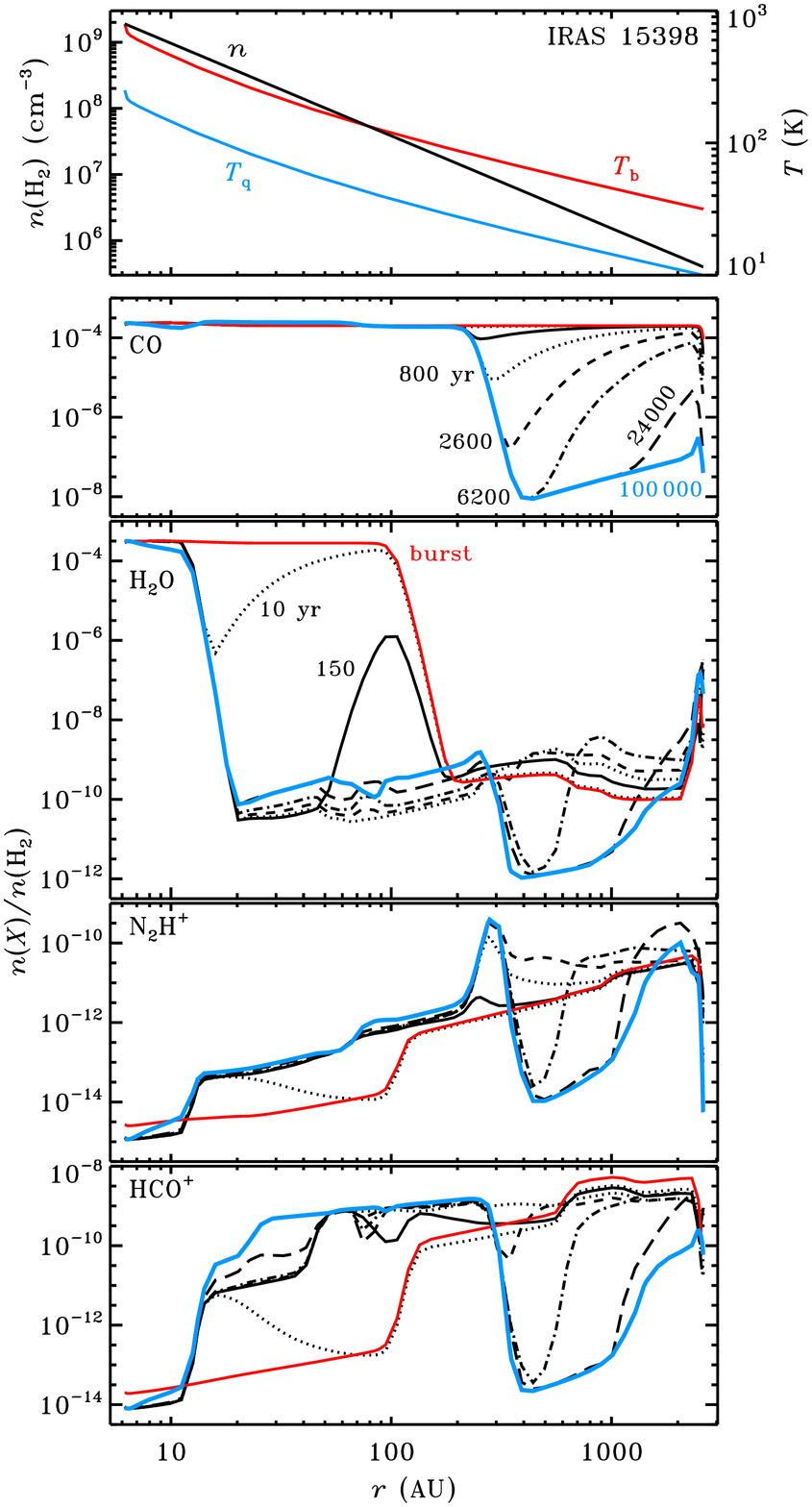}
\caption{\emph{Top panel:} radial density profile in IRAS 15398 (black), along with the temperatures in the quiescent phase (blue) and during the burst (red). \emph{Other panels:} radial abundance profiles of CO, \w, \nnhp, and \hcop. The red curves are during the accretion burst. The black curves are during the quiescent phase, at increasing amounts of time after the burst: 10 yr (dotted), 150 yr (solid), 800 yr (dotted), 2600 yr (short-dashed), 6200 yr (dash-dotted), and 24\,000 yr (long-dashed). The quiescent profiles culminate in the blue curve at \scit{1}{5} yr after the burst.}
\label{fig:abun15398}
\end{figure}

The heating and cooling rates of the dust are fast enough for our purposes that the dust temperature responds instantaneously to a change in luminosity \citep{johnstone13a}. We set the gas temperature equal to the dust temperature at all times. As noted in \vb, the temperature at any point in the envelope is proportional to $L_\ast^{0.25}$ and thus increases by a factor of 3.2 if the star becomes 100 times brighter \citep[see also][]{johnstone13a}. As an example, the top panel of \fig{abun15398} in \sect{res} shows the ``quiescent'' and ``burst'' temperature profiles for IRAS 15398.


\subsection{Chemical network}
\label{sec:chemnet}
The basis of our chemical network is the \textsc{Rate12} release of the UMIST Database for Astrochemistry \citep{mcelroy13a}. In addition to standard neutral-neutral and ion-molecule chemistry, \textsc{Rate12} includes photodissociation and photoionization reactions. These are important only at the outer edges of the envelopes, which we assume are irradiated by the mean interstellar radiation field. The cosmic-ray ionization rate is set to a standard value of \scit{5}{-17} \ps{} \citep{dalgarno06a}. X-rays are not included; they have been detected in one T Tauri FUor \citep{liebhart14a}, but X-ray luminosities in embedded protostars remain highly uncertain and are beyond the scope of this work.

We expand the network with freeze-out and evaporation of all neutral molecules. The freeze-out timescale depends inversely on the density and is typically of similar duration ($\sim$\ten{3}--\ten{4} yr) as the quiescent phase between subsequent accretion bursts, allowing for an abundance pattern out of equilibrium with the stellar luminosity (\citealt{lee07a}; \vb). Indeed, it is the long freeze-out timescale that drives most of the chemical aspects of episodic accretion explored in this work.

Non-thermal desorption via cosmic rays \citep{hasegawa93a} or UV radiation \citep{oberg07a} is included but only plays a minor role in episodic accretion chemistry. Much more important is thermal desorption. According to spectroscopic observations, protostellar CO evaporates at 25--30 K \citep{jorgensen02a,jorgensen04c,jorgensen15a,yildiz13a}. This is at least 5 K higher than predicted from binding energies measured in the laboratory, especially the ``pure ice'' value of 855 K from \citet{bisschop06a} used in \vb\@. Hence, we now adopt a binding energy of 1307 K, measured for CO evaporating from amorphous water ice \citep{noble12a}. Under protostellar conditions, this higher binding energy is consistent with the empirical evaporation temperature of 25--30 K\@. Likewise, we increase the binding energy of \mn{} from 800 K to 1200 K\@. The binding energy for atomic O has long been set to 800 K \citep{watson72a}, but new experiments on water ice surfaces offer conclusive evidence that this old estimate is too low (\citealt{minissale14a}; Minissale et al.\ in prep.). We adopt the current best estimate of 1420 K\@. The binding energies for other molecules are unchanged from \vb, such as 5773 K for \w{} and 2300 K for \cdo{} \citep{fraser01a,noble12a}.

Our network contains two types of grain-surface reactions. The first is the usual formation of \mh{} \citep{black87a}. The second type is simple hydrogenation of C to \meh, N to \amh, and O to \w\@. This happens one H atom at a time at a rate equal to the adsorption rate of H onto the grain surface multiplied by the relative fraction of the reactant molecule in the ice \citep{visser11a}. Conversion of CO ice into \cdo{} ice, as proposed by \citet{kim11a,kim12a}, is not included; see \sect{modobs} for details.

The elemental abundances relative to the total number of H atoms are 0.09 for He, \scit{1.4}{-4} for C, \scit{7.5}{-5} for N, and \scit{3.2}{-4} for O \citep{mcelroy13a}. At the start of the first accretion burst, the chemical composition is set to typical pre-stellar core conditions \citep{maret06a,whittet09a}: hydrogen in atomic H (0.005\%) and \mh{} ($\sim$100\%); carbon in CO (37\%), CO ice (35\%), and \cdo{} ice (28\%); remaining oxygen in \w{} ice; and nitrogen in atomic N (55\%), \mn{} (42\%), and \amh{} ice (3\%). The initial ratio of \cdo{} ice to CO ice is 0.8:1, the average value observed in three dense molecular clouds believed to be representative of the earliest stage of star formation \citep{whittet09a}.

\subsection{Line radiative transfer}
\label{sec:radtran}
In order to compare our model to observations, we simulate molecular line spectra with the one-dimensional radiative transfer code RATRAN \citep{hogerheijde00a}. RATRAN solves for the level populations as function of position, accounting for both collisional and radiative excitation, and then performs ray tracing to synthesize a spectral image. Lastly, the images are convolved to the appropriate telescope beam for each source and transition (\tb{linedata}).

The low-$J$ lines of CO and \hcop{} are optically thick for most sources, so the optically thin isotopologs \ceo{} and \htcop{} are used instead. The isotope ratios are set to 69 for $^{12}$C/$^{13}$C and 557 for $^{16}$O/$^{18}$O \citep{wilson99a}. Collision rates are taken from \citet{flower99a}, \citet{daniel05a}, and \citet{yang10a}, as compiled in the Leiden Atomic and Molecular Database\footnote{\url{http://home.strw.leidenuniv.nl/~moldata}} \citep[LAMDA;][]{schoier05a}.


\section{Effects of accretion bursts on abundance profiles and rotational spectra}
\label{sec:res}
As described in \sect{envmod}, we simulate an accretion burst in all 25 sources by increasing the luminosity to 100 times the observed luminosity for a duration of 100 yr. Starting with the pre-stellar core composition from \sect{chemnet}, we evolve the chemistry during the burst at 78--89 radial points per source. After 100 yr, the burst ends and the temperature instantaneously returns to the quiescent profile. Continuing with the abundances from the end of the burst, we evolve the chemistry for another \ten{5} yr and extract abundances at a number of intermediate time steps.

All sources have the same qualitative abundance profiles and molecular line spectra, so we choose IRAS 15398 for a quantitative discussion. The results for the rest of the sample are presented in \sect{tscales}.

\begin{figure*}[t!]
\centering
\includegraphics[width=\hsize]{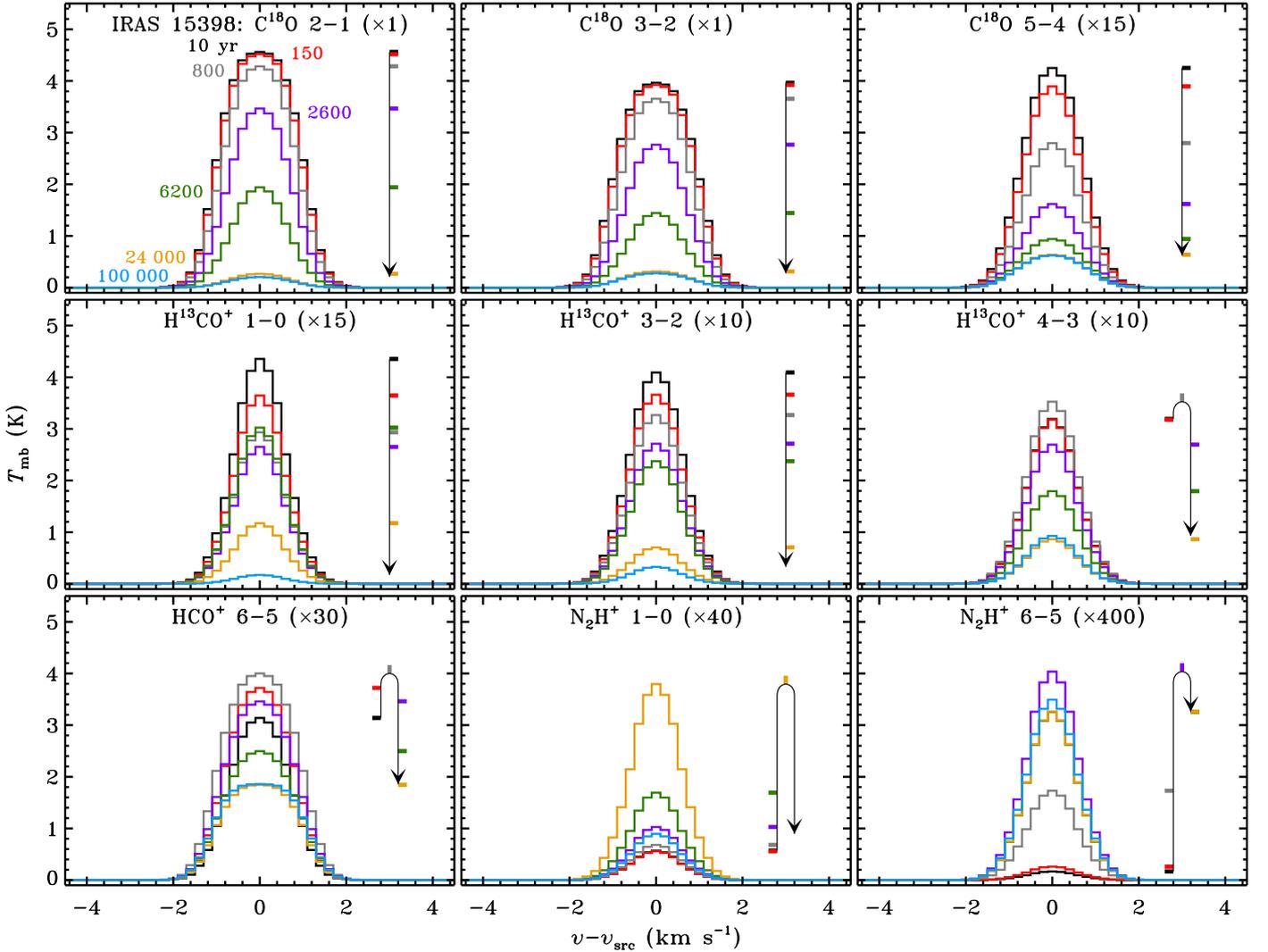}
\caption{Synthetic spectra for the quiescent phase in IRAS 15398, scaled as indicated. The different colors correspond to the same post-burst time steps as in \fig{abun15398}: 10 yr (black), 150 yr (red), 800 yr (gray), 2600 yr (purple), 6200 yr (green), 24\,000 yr (orange), and \scit{1}{5} yr (blue). The arrows indicate how the intensity of each line changes as function of time; the colored flags mark the peak intensities at the listed times. The \nnhp{} 1--0 panel shows the isolated $F_1=0$--1 hyperfine component, whereas the 6--5 panel shows the sum over all unresolved components.}
\label{fig:spec15398}
\end{figure*}


\subsection{Abundance profiles}
\label{sec:abun}
IRAS 15398 currently has a luminosity of 1.6 $L_\odot$ and an envelope mass of 0.5 $M_\odot$ \citep{kristensen12a}. The top panel of \fig{abun15398} plots the temperature profiles for the quiescent phase ($T\q{q}$) and the burst ($T\q{b}$) in IRAS 15398, along with the density profile ($n$). The quiescent temperature decreases from 250 K at the inner edge of the envelope to 10 K at the outer edge. During the burst, the entire curve goes up by a factor of 3.2 (\sect{envmod}).

The other panels of \fig{abun15398} show the radial abundance profiles of CO, \w, \nnhp, and \hcop{} during the burst (red curves) and at a series of time steps after the burst (black and blue). The basic results for CO are the same as reported elsewhere (\citealt{lee07a}; \vb; \citealt{vorobyov13a}). The temperature exceeds 30 K everywhere during the burst and more than 99.9\% of CO is in the gas phase at all radii.\footnote{\citet{lee07a} simulated a lower-luminosity protostar and retained some CO ice in the outer envelope during the burst.} After the burst, CO freezes out beyond 250 AU, where the quiescent temperature lies below 25~K\@. Since the freeze-out rate is proportional to the gas density, depletion is fastest right around the 25 K radius. For the specific density profile in the envelope model of IRAS 15398, it takes 700 yr for the CO abundance to drop by a factor of 10 and 1500 yr for a factor of 100.

The column density of CO ice integrated through the envelope is reduced almost to zero during the burst. Once the envelope cools down, the ice column starts to reform. It reaches 20\% of the original amount after 100 yr and 90\% after 4000 yr. \cdo{} has a higher binding energy than CO and evaporates around 45~K\@. This allows 15\% of the integrated \cdo{} ice column to survive during the simulated burst in IRAS 15398. It takes 600 yr in the quiescent phase to rebuild 90\% of the original \cdo{} ice.

\w{} evaporates out to 100 AU during the burst. Like CO, the excess \w{} refreezes in the quiescent phase, but it does so about ten times faster because of the higher densities in the inner envelope. All \w{} that evaporated during the burst disappears within a few 100 yr in the envelope model of IRAS 15398. On timescales of a few 1000 yr, the freeze-out of CO removes a substantial amount of oxygen from the gas phase. This leads to a drop in the \w{} profile near the 25 K radius at 250 AU\@. The \w{} abundance increases again towards the outer envelope because of photodesorption. The integrated column density of \w{} ice drops by 65\% during the burst; afterwards, it takes only $\sim$10 yr to rebuild the column to 90\% of the equilibrium value.

The chemistry of \nnhp{} is strongly tied to that of CO and \w, because both act as important destruction channels. The depletion of CO and \w{} in the quiescent phase thus allows \nnhp{} to increase in abundance at almost all radii for the first few 1000 yr after the burst. This type of anti-correlation between CO and \nnhp{} is well known from observations of molecular clouds, embedded protostars, and circumstellar disks \citep{bergin02a,jorgensen04d,qi13a} and was also noted by \citet{lee07a}. On timescales of more than a few 1000 yr, freeze-out of \mn{} near 400 AU causes a decrease in the \nnhp{} abundance.

Lastly, the abundance of \hcop{} undergoes a seesaw motion around the 25 K radius. CO is the primary parent species of \hcop{} and \w{} is one of the dominant destroyers. Freeze-out of \w{} allows more \hcop{} to be formed in the inner envelope, while freeze-out of CO reduces the amount of \hcop{} in the outer envelope on longer timescales. The correlation between the protostellar abundances of CO and \hcop{} is another well-established observational result \citep{jorgensen04c}.

The chemical evolution in the other 24 envelope models is qualitatively the same as in IRAS 15398, but the timescales on which the aforementioned changes occur differ depending on the density profiles. IRAS 15398 has relatively high densities in the \w{} and CO freeze-out zones compared to the rest of the sample. As a consequence, freeze-out in the other sources is typically slower by up to two orders of magnitude and the chemical effects of the accretion burst take longer to subside.

In summary, the chemical composition of a protostellar envelope changes substantially during an accretion burst. These changes last for long after the system returns to quiescence:
\begin{itemize} \itemsep1pt
\vspace{-0.4em}
\item[$\bullet$] enhanced CO in the outer envelope for a few \ten{3}--\ten{4} yr;
\item[$\bullet$] reduced CO ice and \cdo{} ice in the outer envelope for a few \ten{3}--\ten{4} yr;
\item[$\bullet$] enhanced \w{} in the inner envelope for a few \ten{2}--\ten{3} yr;
\item[$\bullet$] reduced \nnhp{} in the outer envelope for a few \ten{3}--\ten{4} yr;
\item[$\bullet$] seesaw pattern for \hcop: enhanced in the inner envelope, reduced in the outer envelope for a few \ten{3}--\ten{4} yr.
\vspace{-0.4em}
\end{itemize}
The exact timescales in a given source depend on the density profile. The next section explores the consequences of the abundance changes on the rotational emission lines.

\begin{figure}[t!]
\centering
\includegraphics[width=\hsize]{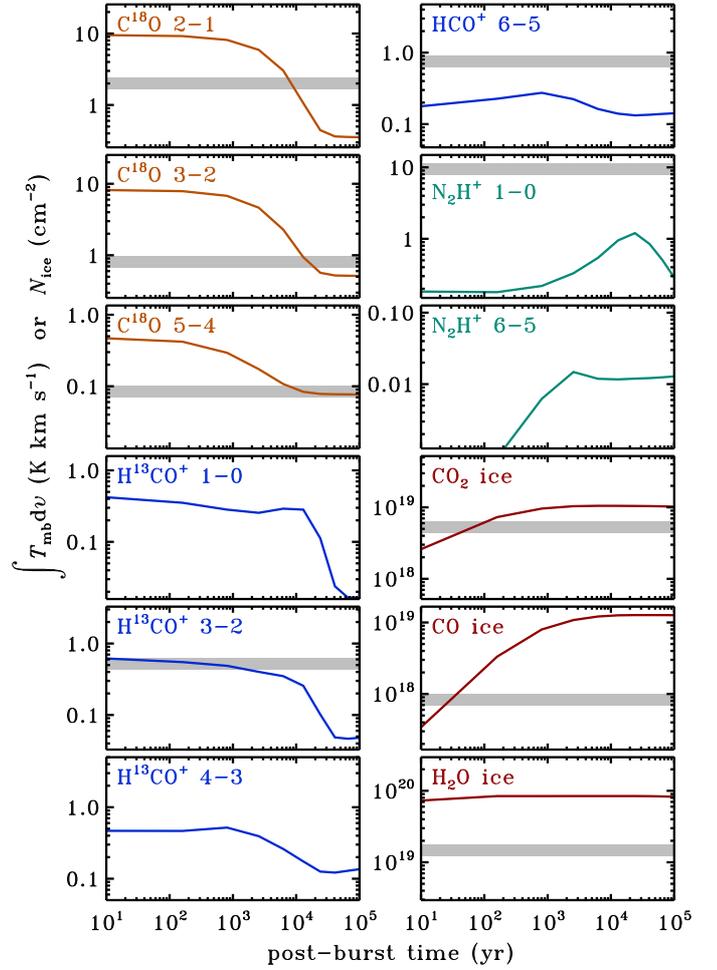}
\caption{Simulated integrated line intensities and ice column densities in IRAS 15398 as function of time during the quiescent phase. The \nnhp{} $J=1$--0 and 6--5 intensities are summed over all HF components. Where available, the gray bars show the observed intensities and columns from Tables \ref{tb:linedata} and \ref{tb:icedata} with 20\% uncertainties.}
\label{fig:mod-obs-1}
\end{figure}


\subsection{Rotational spectra and ice column densities}
\label{sec:spec}
The abundance changes caused by an accretion burst persist for long after the burst has ended. To what extent can the chemical effects of a burst still be observed once the protostar has returned to quiescence? \figg{spec15398} shows a set of nine single-dish spectra corresponding to the abundance profiles at the quiescent time steps from \fig{abun15398}. All spectra are first convolved with the appropriate telescope beam (\tb{linedata}) and then continuum-subtracted. The arrow in each panel indicates schematically how the peak intensity changes with time: continuous decrease, continuous increase, or an increase followed by a decrease.

For a more quantitative view of the temporal behavior, the spectra from \fig{spec15398} are integrated and plotted as function of time in \fig{mod-obs-1}. Also shown here are the simulated column densities of \cdo, CO, and \w{} ice. Where available, \fig{mod-obs-1} includes the observed intensities and column densities from \sect{obs} as gray bars with 20\% error margins. We compare the model results to the observations in \sect{modobs}; the remainder of the current section discusses the model trends on their own.

During the quiescent phase, the simulated line intensities change predictably based on the abundance variations. The freeze-out of CO leads to a monotonic decrease in the optically thin \ceo{} lines. \htcop{} 1--0 and 3--2 are dominated by the cold outer envelope and their intensities also decrease monotonically as CO and \hcop{} are depleted. \htcop{} 4--3 and \hcop{} 6--5 originate closer to the star and pick up on the seesaw abundance pattern from \fig{abun15398}: the line intensities show an initial rise before depletion in the outer envelope sets in and the intensities go down.

Because of the abundance anticorrelation between CO and \nnhp, the line intensities of \ceo{} and \nnhp{} evolve largely in opposite directions. As CO and \w{} freeze out after the burst, \nnhp{} becomes more abundant and its 1--0 and 6--5 lines gain intensity. The 1--0 intensity decreases again on timescales of more than $\sim$24\,000 yr after the burst. This is due to the depletion of atomic N and the resulting loss of \nnhp{} in the cold outer envelope. The 6--5 line is not very sensitive to the \nnhp{} abundance around 1000 AU; its intensity levels off at late times, but does not turn over.

The total observable column densities of \cdo{} and CO ice increase by factors of 3 and 20 on timescales of \ten{3}--\ten{4} yr after the burst. The column of \w{} ice is only marginally affected by changes in luminosity and remains constant at all times.


\section{Line ratios as a diagnostic of episodic accretion timescales}
\label{sec:tscales}
One of the challenges in episodic accretion is to understand the magnitude--frequency distribution: how often do burst of a certain magnitude or intensity occur? The chemical effects explored in this work can help constrain the frequency of very strong bursts, where the accretion rate increases by at least a factor of 100. In the quiescent phase following such a burst, several observables vary monotonically with time. In principle, any of these can be used as a chronometer of when the most recent burst occurred. For a statistical sample of protostars, it then becomes possible to determine the average burst frequency.

With the current set of spherical models, inferring when the most recent burst occurred in any given source will be a rather crude method with uncertainties at the order-of-magnitude level. The following discussion should therefore be seen as an attempt to explore certain methodologies, rather than an attempt to derive firm numbers. For an alternative approach based on spatially resolved observations of \ceo, see \citet{jorgensen15a}.


\subsection{Models vs.\ observations}
\label{sec:modobs}
\figg{mod-obs-1} highlights both the successes and the shortcomings of our model for IRAS 15398. Six of the observables are reproduced at some point during the quiescent phase, but the times at which the matches occur range over more than two orders of magnitude: from $\sim$30 yr for CO ice to $\sim$\ten{4} yr for \ceo{} 2--1. Furthermore, \w{} ice is overproduced at all times, and \hcop{} 6--5 and \nnhp{} 1--0 are underproduced.

The overproduction of \w{} ice may be due to a lack of grain-surface reactions in our chemical network, which tend to drain oxygen out of \w{} and convert it to other species \citep{schmalzl14a}. \nnhp{} 1--0 probably suffers from our choice of a spherically symmetric envelope model. In spatially resolved observations, the morphology of \nnhp{} 1--0 often deviates significantly from spherical symmetry \citep{jorgensen04d,tobin11a}. This can result in enhanced emission in one direction, e.g. from interactions between the outflow and the dense envelope, which our spherical models do not reproduce. The underproduction of \hcop{} 6--5 appears to be an excitation effect: this line is enhanced along the outflow cavity walls due to direct ultraviolet heating, similar to what is seen in $^{12}$CO and $^{13}$CO 6--5 \citep{spaans95a,yildiz12a}.

\citet{kim11a,kim12a} argued that some fraction of CO ice needs to be converted into \cdo{} ice during each quiescent phase in between subsequent bursts. However, that conclusion may be an artifact of the low initial \cdo{} abundance in their model. Our model starts with the average CO and \cdo{} ice abundances observed towards dense cores \citep{whittet09a}. As a result, it does not require any conversion of CO into \cdo{} ice to reproduce the observed column densities in IRAS 15398.

\begin{figure*}[t!]
\centering
\includegraphics[width=\hsize]{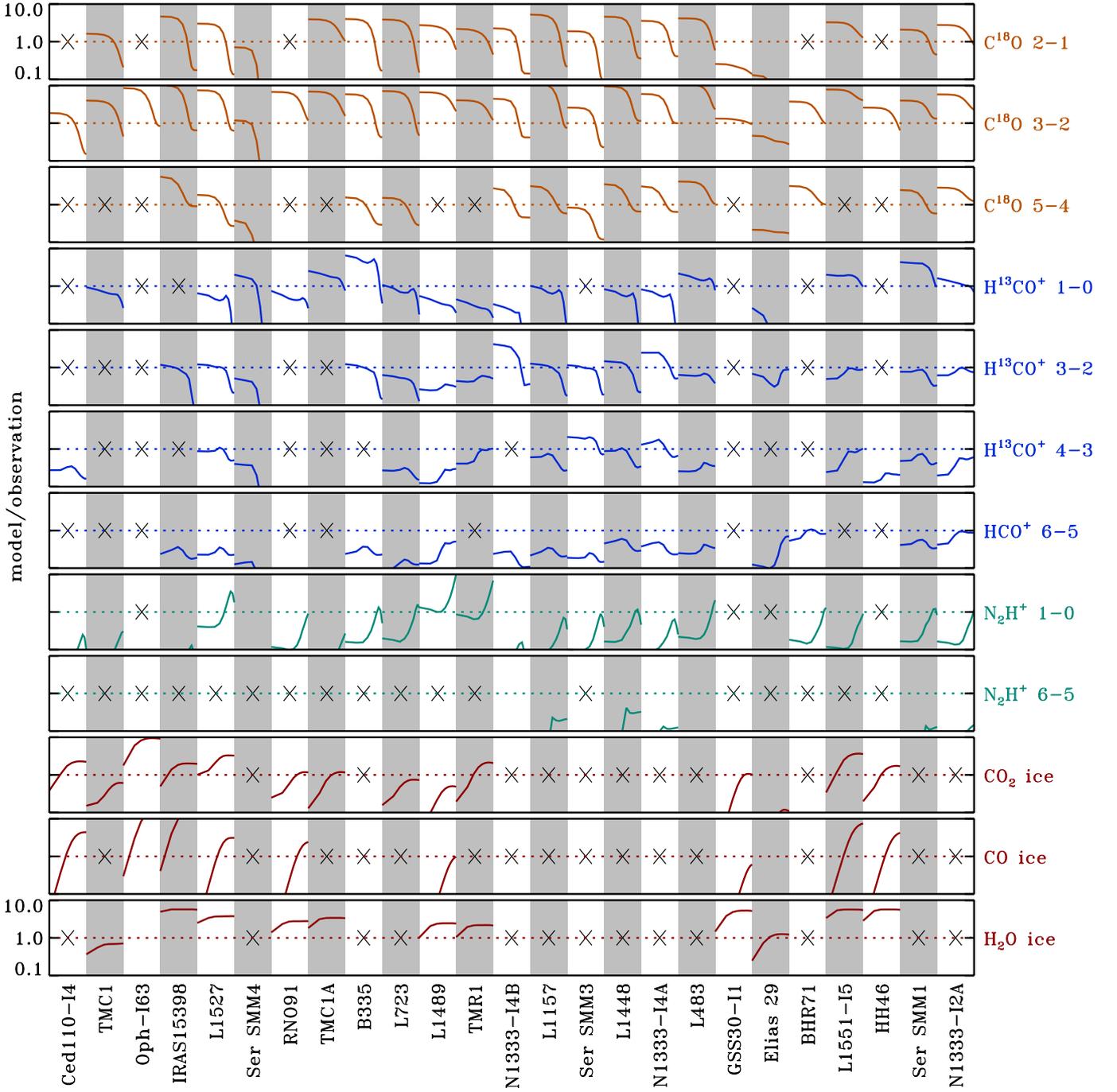}
\caption{Comparison between model predictions and observations. This plot is similar to \fig{mod-obs-1}, except that the model values are now divided by the observed quantities. A ratio of unity means the model is in exact agreement with observations. If the ratio is larger than unity, the model overpredicts the observation, and vice versa. The horizontal scale for each source runs logarithmically from 10 to \ten{5} yr after the burst. The observable quantity in the bottom three panels is the column density of ice along a pencil beam to the central star. In the other panels, it is the integrated intensity of various rotational lines. The \nnhp{} $J=1$--0 and 6--5 intensities are summed over all HF components. The sources are ordered from left to right by increasing luminosity.}
\label{fig:mod-obs}
\end{figure*}

The comparison between model predictions and observations is expanded to all 25 protostars in \fig{mod-obs}. The trends identified in Figs.\ \ref{fig:spec15398} and \ref{fig:mod-obs-1} for IRAS 15398 generally hold for the full sample: all \ceo{} lines become weaker as function of time and usually match the observed intensities at some point. \htcop{} 1--0 and 3--2 also decrease monotonically and are in reasonable agreement with the observations. \htcop{} 4--3 and \hcop{} 6--5 reflect the seesaw abundance pattern of \hcop{} by first increasing and then decreasing in strength. Both lines are too weak in the models by a factor of 2--10. The \nnhp{} 1--0 line gains intensity during the quiescent phase in all sources and turns over at late times in about a third of the sample. \nnhp{} 6--5 initially increases and then levels off. Both \nnhp{} lines tend to be underproduced relative to the observed intensities. The three ice column densities always increase with time after the burst. \cdo{} ice and CO ice usually match the observations, but \w{} ice is consistently too abundant in the models.

In order to construct a consistent set of models, we treated all sources as if they are currently in the quiescent phase (\sect{envmod}). This is undoubtedly a false assumption -- especially for the brighter sources in our target list -- and adds another source of discrepancies between models and observations. Future studies of individual sources or large samples will benefit from independent constraints on the accretion rates, e.g. from hydrogen Br$\gamma$ \citep{connelley10a}. If the stellar mass is known from measurements of a rotationally supported disk \citep{tobin12a}, the accretion rate can also be computed from the relationship between $L\q{bol}$, $M_\ast$, and $\dot{M}$ \citep{adams86a}.


\subsection{Line ratios}
\label{sec:ratios}
The absolute line intensities in \fig{mod-obs-1} are sensitive to uncertainties in the source distance and in various model parameters like the envelope mass and the temperature profile. A better approach is to take the ratio between two rotational lines whose intensities change in opposite directions after the burst. This effectively acts as a self-normalization against distance and source model, in particular for line pairs at similar excitation levels.

\tb{coratios} lists thirteen observable ratios as candidate chronometers, chosen in part for their sensitivity to the burst, and in part for being easily observable with modern sub-millimeter facilities. The first four ratios are the low-$J$ line pairs of \ceo{} 2--1, \ceo{} 3--2, \htcop{} 1--0, and \htcop{} 3--2 over \nnhp{} 1--0. The fifth one is the high-$J$ pair of \ceo{} 5--4 over \nnhp{} 6--5. The remaining eight are the ratios between the intensities of \ceo{} 2--1, \ceo{} 3--2, \htcop{} 1--0, and \htcop{} 3--2 on the one hand and the column densities of \cdo{} or CO ice on the other hand.

\begin{table}
\caption{Candidate observable ratios to be used as chronometers of episodic accretion timescales.}
\label{tb:coratios}
\centering
\begin{tabular}{c}
\hline\hline
\ceo{} 2--1 / \nnhp{} 1--0 \hspace{1em} \ceo{} 3--2 / \nnhp{} 1--0\rule{0pt}{1.0em} \\
\htcop{} 1--0 / \nnhp{} 1--0 \hspace{1em} \htcop{} 3--2 / \nnhp{} 1--0 \\
\ceo{} 5--4 / \nnhp{} 6--5 \\
\ceo{} 2--1 / \cdo{} ice \hspace{1em} \ceo{} 2--1 / CO ice \\
\ceo{} 3--2 / \cdo{} ice \hspace{1em} \ceo{} 3--2 / CO ice \\
\htcop{} 1--0 / \cdo{} ice \hspace{1em} \htcop{} 1--0 / CO ice \\
\htcop{} 3--2 / \cdo{} ice \hspace{1em} \htcop{} 3--2 / CO ice \\
\hline
\end{tabular}
\end{table}

\begin{figure*}[t!]
\centering
\includegraphics[width=\hsize]{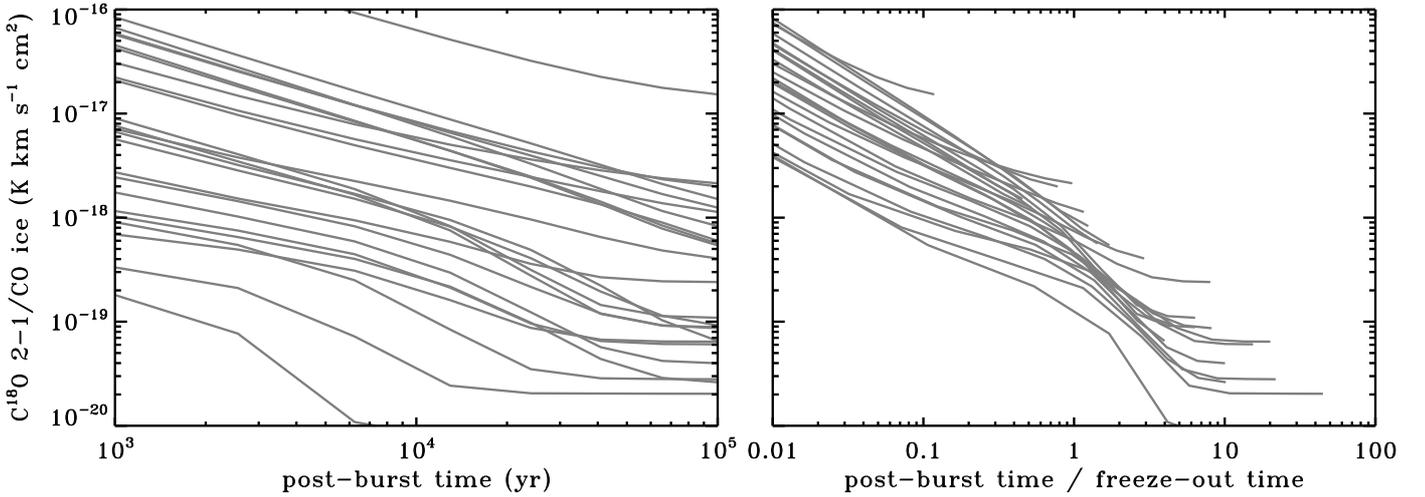}
\caption{\emph{Left:} ratio between the integrated intensity of \ceo{} 2--1 and the column density of CO ice as function of time during the quiescent phase. Each gray curve corresponds to one source. \emph{Right:} same, but with the time axis normalized by the characteristic freeze-out timescale for each source.}
\label{fig:lratio}
\end{figure*}

As an example, the left panel of \fig{lratio} shows how the ratio between the integrated intensity of \ceo{} 2--1 and the column density of CO ice evolves as function of time after the burst for the full source sample. The ratio plots for the other 12 pairs of observables in \tb{coratios} look qualitatively the same. Each curve represents one source model. The ratios decreases with time for all sources, but the curves are scattered across more than four orders of magnitude and are of little diagnostic value in this form.

The scatter in \fig{lratio} is due to the range of envelope parameters encountered across the sample. CO freezes out at the same temperature in each source ($<25$ K), but because of all the different envelope parameters, it does \emph{not} do so at the same density. In high-density envelopes, the freeze-out timescale is shorter and the CO gas/ice abundance ratio decreases more rapidly than in lower-density sources. Likewise, the \ceo/\nnhp{} and \htcop/\nnhp{} line intensity ratios change at different rates for different sources.

The scatter in the left panel of \fig{lratio} can be reduced by expressing the time axis for each source in units of its characteristic freeze-out timescale (\vb, erratum):
\begin{equation}
\label{eq:taufr}
\tau\q{fr} = 1\times10^4\,{\rm yr}\,\sqrt{\frac{10\,{\rm K}}{T\q{fr}}} \frac{10^6\,{\rm cm}^{-3}}{n\q{fr}}\,.
\end{equation}
The characteristic freeze-out density $n\q{fr}$ is defined as the density at which the temperature reaches $T\q{fr}$ in the quiescent phase. We tested $T\q{fr}$ from 10 to 40 K and found the greatest reduction in scatter for 15 K, where the bulk of the low-$J$ \ceo{} emission originates. At 15 K, $n\q{fr}$ ranges from \scit{3.5}{3} \pcc{} in Elias 29 to \scit{5.5}{6} \pcc{} in Ser SMM4. The corresponding freeze-out timescales vary from 2.3 Myr to 1500 yr.

The time-normalized \ceo{} 2--1/CO ice ratios appear as the gray curves in the right panel of \fig{lratio}. The sources are still scattered across an order of magnitude, but all follow the same trend and an average relationship between normalized post-burst time and gas/ice observable ratio can easily be drawn. Our source sample is probably diverse enough (albeit not unbiased) that we can use this average relationship also for other samples where detailed envelope models are not readily available.


\subsection{Future prospects}
\label{sec:future}
Using all the observable ratios from \tb{coratios}, we can estimate when the most recent accretion burst occurred in each of our sources. However, the results are currently too scattered and the uncertainties too large to produce reliable numbers worth tabulating. For example, \citet{jorgensen13a} concluded from spatially resolved \htcop{} 4--3 that IRAS 15398 experienced a burst between 100 and 1000 yr ago. We find the same timescale from the \ceo{} gas/\cdo{} ice ratios, but \ceo{} gas/CO ice suggests a longer timescale of $\sim$6000 yr, whereas \htcop{} 3--2/\nnhp{} 1--0 yields a value as long as \scit{5}{4} yr. Furthermore, the observed \ceo/\nnhp{} line ratios do not match any timescale from the model. The situation is no better for most of the other sources. This lack of consistency between different tracers was already evident in Figs.\ \ref{fig:mod-obs-1} and \ref{fig:mod-obs}, where some curves match the observations at very different post-burst times (\sect{modobs}).

The most likely reason for these discrepancies is the assumption of spherical symmetry for the envelope models (\sect{modobs}). However, even if the spherical models do not produce reliable timescales, they are still useful as a basis for follow-up studies. Spatially resolved observations offer many advantages over single-dish data (as explored for e.g.\ \ceo{} 2--1 by \citealt{jorgensen15a}) and can be carried out for statistically significant source samples with current facilities. We will expand the models beyond spherical symmetry and calibrate them against existing interferometric data to identify the most reliable tracers of episodic accretion timescales. These tracers can then be used to derive timescales for many tens or hundreds of protostars and compared against results from other methods. Working with such large samples also reduces the problem that any given source may be an outlier.


\section{Conclusions}
\label{sec:conc}
This paper simulates the chemical effects of episodic accretion in a diverse sample of 25 protostellar envelope models. The intensity of the accretion bursts is set to 100 times the quiescent luminosity, enough to evaporate CO ice all the way to the outer edge of the envelope in each source (\sect{abun}). The \w{} snowline expands in size by a factor of $\sim$10 during the burst. The evaporation of CO and \w{} leads to the destruction of \nnhp{} throughout the envelope. \hcop{} is enhanced outside of the \w{} snowline and destroyed inside of it.

When the burst ends, the temperatures quickly return to normal. The species that evaporated during the burst start to refreeze onto the cold dust, but this is a relatively slow process. \w{} remains enhanced for \ten{2}--\ten{3} yr and CO for an order of magnitude longer. The abundances of \nnhp{} and \hcop{} are reset on similarly long timescales.

The abundance changes resulting from an accretion burst also have a substantial effect on various molecular lines, as quantified by radiative transfer simulations at a series of time steps (\sect{spec}). \ceo{} 2--1, 3--2, and 5--4 and \htcop{} 1--0 and \mbox{3--2} are enhanced during the burst and decay monotonically after the burst ends. \htcop{} 4--3, \hcop{} 6--5, and \nnhp{} 1--0 first increase after the burst and then decrease. The column densities of CO and \cdo{} ice increase monotonically after the burst.

The simulated line intensities and ice column densities are compared to single-dish observations gathered from the literature and to previously unpublished spectra of \hcop{} and \nnhp{} 6--5 from the Herschel Space Observatory. Several line ratios are identified that can be used, in principle, as a chronometer of when the most recent strong burst occurred in each source (\sect{tscales}). In practice, however, the spherical source models are not accurate enough to derive reliable timescales from single-dish data. Interferometric observations are therefore recommended for follow-up work. In order to tackle a statistically significant source sample, the models need to be calibrated against spatially resolved observations to identify robust tracers of episodic accretion timescales.


\begin{acknowledgements}
This work was supported by the National Science Foundation under grant 1008800 and by a Postdoctoral Fellowship from the European Southern Observatory. JKJ was supported by a Lundbeck Foundation Junior Group Leader Fellowship and by the Centre for Star and Planet Formation, funded by the Danish National Research Foundation. This research made use of NASA's Astrophysics Data System and of the SIMBAD database, operated at CDS, Strasbourg, France.
\end{acknowledgements}



\appendix
\section{Details of observations}
\label{sec:obsdetails}
The spectroscopic observations used in this work are summarized in Tables \ref{tb:linedata} and \ref{tb:icedata}. The following subsections offer more details on how the data were obtained and reduced. The \hcop{} and \nnhp{} 6--5 spectra are published here for the first time; all other observations were gathered from the literature. Typical calibration uncertainties are between 10\% and 25\%.

\subsection{C\hspace{0.05em}$^\mathit{1}$\hspace{-0.10em}$^\mathit{8}$\hspace{-0.10em}O}
\label{sec:obsc18o}
\citet{yildiz13a} compiled all available \ceo{} data from $J=2$--1 up to 10--9, with one omission: an integrated intensity of 11.0 \kkms{} in a $22.9''$ beam for the 2--1 line in Ser SMM4 \citep{hogerheijde99a}. The 6--5 and higher lines are detected in only a handful of sources and are therefore of limited use as an observational diagnostic. The 2--1, 3--2, and 5--4 lines have detection rates of 77\%, 100\%, and 58\% and are well suited for our purpose. Furthermore, with upper-level energies of 16, 32, and 79 K, these three lines together are a good probe of the abundance jump at the CO sublimation front \citep{yildiz13a}.


\subsection{HCO\hspace{0.05em}$^\mathit{+}$}
\label{sec:obshco+}
The low-$J$ lines of \hcop{} are often optically thick, so we only consider the 6--5 line at 535 GHz ($\eup/k=90$ K). This transition was observed with Herschel-HIFI as part of WISH and the related open-time program ``Water in low-mass protostars: the William Herschel line legacy'' (WILL; proposal code OT2\_evandish\_4; Mottram et al.\ in prep.). Five of the spectra were presented by \citet{sanjose14a} and \citet{benz15a}. The other eleven \hcop{} lines from this dataset are as yet unpublished.

\tb{hobs} lists the observing dates and identification numbers (ObsIDs), including several duplicate or triplicate observations taken in spectral setups of other lines. Five of these come from the open-time program ``Searching for the onset of energetic particle irradiation in Class 0 protostars'' (OT2\_cceccare\_4).

\begin{table}
\caption{Herschel observations of \hcop{} and \nnhp.}
\label{tb:hobs}
\centering
\begin{tabular}{lccc}
\hline\hline
Source & ObsID & Date & Ref.\tablefootmark{a}\rule{0pt}{0.9em} \\
\hline
\multicolumn{4}{c}{\hcop{} 6--5 at 535.062 GHz}\rule{0pt}{0.9em}\\
\hline
L1448 MM & 1342203186 & 2010-08-18 & 1\rule{0pt}{0.9em} \\
NGC1333 IRAS2A & 1342192206 & 2010-03-14 & 1 \\
               & 1342202024 & 2010-08-01 & 1 \\
               & 1342248912 & 2012-07-31 & 2 \\
NGC1333 IRAS4A & 1342192207 & 2010-03-15 & 1 \\
               & 1342202023 & 2010-08-01 & 1 \\
               & 1342248914 & 2012-07-31 & 2 \\
NGC1333 IRAS4B & 1342192208 & 2010-03-15 & 1 \\
               & 1342202022 & 2010-08-01 & 1 \\
               & 1342202033 & 2010-08-02 & 1 \\
L1527 & 1342203188 & 2010-08-18 & 1 \\
      & 1342250193 & 2012-08-24 & 2 \\
BHR71 & 1342200755 & 2010-07-07 & 1 \\
IRAS 15398 & 1342266008 & 2013-03-05 & 3 \\
L483 MM & 1342207582 & 2010-10-28 & 1 \\
Ser SMM1 & 1342194463 & 2010-04-11 & 1 \\
         & 1342207581 & 2010-10-28 & 1 \\
         & 1342251637 & 2012-09-29 & 2 \\
Ser SMM4 & 1342194464 & 2010-04-11 & 1 \\
Ser SMM3 & 1342207580 & 2010-10-28 & 1 \\
L723 MM & 1342219172 & 2011-04-21 & 1 \\
B335 & 1342219182 & 2011-04-21 & 1 \\
L1157 & 1342199077 & 2010-06-23 & 1 \\
      & 1342246462 & 2012-05-17 & 2 \\
L1489 & 1342203187 & 2010-08-18 & 1 \\
Elias 29 & 1342266143 & 2013-03-07 & 1 \\
\hline
\multicolumn{4}{c}{\nnhp{} 6--5 at 558.967 GHz\rule{0pt}{0.9em}}\\
\hline
L1448 MM & 1342238644 & 2012-02-03 & 4\rule{0pt}{0.9em} \\
NGC1333 IRAS2A & 1342225935 & 2011-08-09 & 4 \\
               & 1342248913 & 2012-07-31 & 2 \\
NGC1333 IRAS4A & 1342238646 & 2012-02-03 & 4 \\
               & 1342248915 & 2012-07-31 & 2 \\
NGC1333 IRAS4B & 1342238645 & 2012-02-03 & 4 \\
L1527 & 1342250194 & 2012-08-24 & 2 \\
L483 MM & 1342244045 & 2012-04-10 & 4 \\
Ser SMM1 & 1342230370 & 2011-10-08 & 4 \\
         & 1342251636 & 2012-09-29 & 2 \\
L1157 & 1342246461 & 2012-05-17 & 2 \\
\hline
\end{tabular}
\tablefoot{References: (1) WISH: \citealt{sanjose14a}, \citealt{benz15a}; (2) OT2\_cceccare\_4; (3) OT2\_evandish\_4: Mottram et al.\ in prep.; (4) OT1\_philybla\_1.}
\end{table}

All observations were taken in double beam switch mode. We use the spectra recorded with the wide-band spectrometer (WBS; 1.1 MHz resolution) and perform basic standard processing in HIPE v10.0.0 \citep{ott10a}, followed by further analysis in CLASS\footnote{\url{http://www.iram.fr/IRAMFR/GILDAS}}. The H and V polarization spectra were averaged after individual inspection and the intensities were converted to main-beam temperature scale through a main-beam efficiency of 0.72 \citep{roelfsema12a}. Where available, the multiple epochs of data all match to within the calibration uncertainty of 10\% and are averaged to reduce the noise. This results in final rms noise levels of 3--7 mK in 0.28 km \ps{} bins. The final data reduction step was to subtract a linear baseline.

In all but two sources, the \hcop{} 6--5 spectrum shows two emission features: a broad component with a full width at half maximum (FWHM) of 4--14 km \ps{} and a narrow component of 1--2 km \ps{}. These components are also detected in various CO lines and correspond to the outflow (broad) and the quiescent envelope \citep[narrow;][]{yildiz13a,sanjose14a}. The two exceptions in our sample are L723 and B335, which only have the narrow component.

\begin{figure}[t!]
\centering
\includegraphics[width=\hsize]{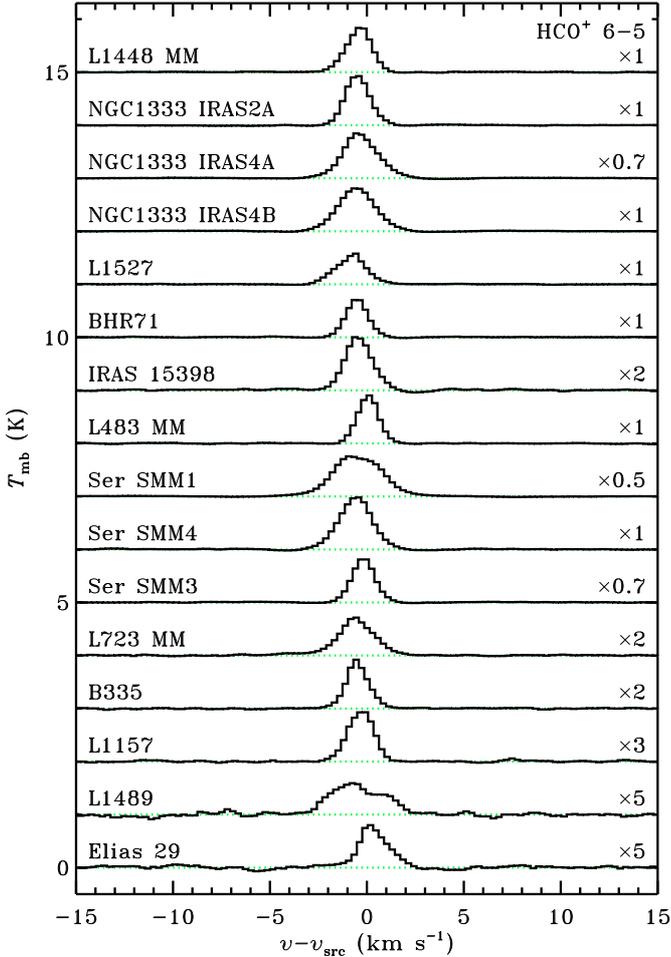}
\caption{Spectra of \hcop{} 6--5, corrected for source velocity and scaled as indicated. If a broad emission component from the outflow was present (in all sources but L723 and B335), it was fitted with a Gaussian profile and subtracted from the spectrum to leave only the narrow emission from the quiescent envelope.}
\label{fig:hco+spec}
\end{figure}

Since we are interested exclusively in the quiescent envelope, we fit two Gaussian profiles to each \hcop{} spectrum (except L723 and B335) and subtract the outflow emission. \figg{hco+spec} shows the resulting spectra, corrected for the source velocities from \citet{yildiz12a}. \tb{linedata} lists the integrated intensities for the full sample.

\subsection{H\hspace{0.05em}$^\mathit{1}$\hspace{-0.10em}$^\mathit{3}$\hspace{-0.15em}CO\hspace{0.05em}$^\mathit{+}$}
\label{sec:obsh13co+}
Complementing the \hcop{} 6--5 line are the 1--0, 3--2, and 4--3 lines of optically thin \htcop. Their upper-level energies are 4, 25, and 42 K, again offering a good probe of chemical changes induced by the evaporation or freeze-out of CO\@.

The integrated intensities compiled in \tb{linedata} originate mostly from the ground-based surveys of \citet{hogerheijde99a} and \citet{jorgensen04c}, with other references as indicated. Some sources have been targeted with multiple telescopes, in which case we choose the observation with the smallest single-dish beam.


\subsection{N\hspace{0.05em}$_\mathit{2}$\hspace{-0.05em}H\hspace{0.05em}$^\mathit{+}$}
\label{sec:obsn2h+}
The final two molecular lines are the $J=1$--0 and 6--5 transitions of \nnhp{} ($\eup/k=4$ and 94 K), both of which are prone to hyperfine splitting \citep{caselli95a}. The quantum numbers for the HF levels are $F_1$ and $F$. The 1--0 transition breaks up into one isolated line ($F_1=0$--1) and two clusters of three lines ($F_1=1$--1 and 2--1), which are readily resolved with heterodyne spectrographs. The 6--5 transition is dominated by nine HF components within 0.1 km \ps{} and therefore appears as a single line in astronomical observations.

Most of the $J=1$--0 intensities in \tb{linedata} are taken from \citet{jorgensen04c} and \citet{emprechtinger09a}, who provided integrated intensities summed over all HF components. Additional data are taken from \citet{mardones97a}, who only showed the spectra of the isolated component and did not tabulate any intensities. We measure the integrated intensities from their Figures 1 and 2 and multiply by a factor of 9 (based on the Einstein A coefficients and level degeneracies) to approximate the intensity for the full 1--0 band.

\begin{figure}[t!]
\centering
\includegraphics[width=\hsize]{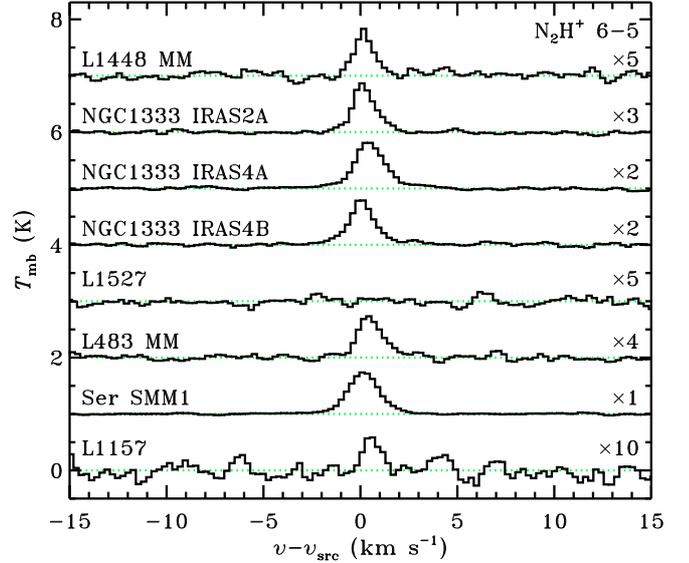}
\caption{Spectra of \nnhp{} 6--5,  corrected for source velocity and scaled as indicated.}
\label{fig:n2h+spec}
\end{figure}

The 6--5 lines were observed primarily as part of the Herschel open-time program ``The chemistry of nitrogen in dark clouds'' (OT1\_philybla\_1), with two additional sources (L1527 and L1157) covered by OT2\_cceccare\_4. Both programs targeted \nnhp{} 6--5 in NGC1333 IRAS2A and 4A and Ser SMM1. None of these spectra have been published, so we follow the same procedure as outlined for \hcop{} 6--5 in \sect{obshco+}. \tb{hobs} lists the observing dates and ID codes and \fig{n2h+spec} shows the reduced spectra. The rms noise is 10--14 mK in 0.27 km \ps{} bins. \nnhp{} 6--5 is not detected in L1527. All other sources show a single narrow emission feature with an FWHM of 1--2 km \ps{} and no contribution from the outflow. \tb{linedata} lists the integrated intensities for the full sample, including the upper limit for L1527.

\subsection{Ices}
\label{sec:obsices}
The ice column densities in \tb{icedata} are based on published mid-infrared observations from the ground and from space. The conversion from spectrum to column density requires an absorption band strength measured in the laboratory for the appropriate pure or mixed ices \citep{gerakines95a}. Most papers on protostellar ices use the same band strengths and we adopt the column densities as published. The two exceptions are the \cdo{} columns of \citet{zasowski09a} and the CO column of \citet{tielens91a}, which we scale by a common factor of 1.4 to be consistent with other studies.


\begin{thebibliography}{98}
\expandafter\ifx\csname natexlab\endcsname\relax\def\natexlab#1{#1}\fi

\bibitem[{{{\'A}brah{\'a}m} {et~al.}(2009){{\'A}brah{\'a}m}, {Juh{\'a}sz},
  {Dullemond}, {K{\'o}sp{\'a}l}, {van Boekel}, {Bouwman}, {Henning},
  {Mo{\'o}r}, {Mosoni}, {Sicilia-Aguilar}, \& {Sipos}}]{abraham09a}
{{\'A}brah{\'a}m}, P., {Juh{\'a}sz}, A., {Dullemond}, C.~P., {et~al.} 2009,
  \nat, 459, 224

\bibitem[{{Adams} \& {Shu}(1986)}]{adams86a}
{Adams}, F.~C. \& {Shu}, F.~H. 1986, \apj, 308, 836

\bibitem[{{Aikawa} {et~al.}(2012){Aikawa}, {Kamuro}, {Sakon}, {Itoh}, {Terada},
  {Noble}, {Pontoppidan}, {Fraser}, {Tamura}, {Kandori}, {Kawamura}, \&
  {Ueno}}]{aikawa12b}
{Aikawa}, Y., {Kamuro}, D., {Sakon}, I., {et~al.} 2012, \aap, 538, A57

\bibitem[{{Arce} {et~al.}(2013){Arce}, {Mardones}, {Corder}, {Garay},
  {Noriega-Crespo}, \& {Raga}}]{arce13a}
{Arce}, H.~G., {Mardones}, D., {Corder}, S.~A., {et~al.} 2013, \apj, 774, 39

\bibitem[{{Audard} {et~al.}(2014){Audard}, {{\'A}brah{\'a}m}, {Dunham},
  {Green}, {Grosso}, {Hamaguchi}, {Kastner}, {K{\'o}sp{\'a}l}, {Lodato},
  {Romanova}, {Skinner}, {Vorobyov}, \& {Zhu}}]{audard14a}
{Audard}, M., {{\'A}brah{\'a}m}, P., {Dunham}, M.~M., {et~al.} 2014, Protostars
  and Planets VI, 387

\bibitem[{{Banzatti} {et~al.}(2012){Banzatti}, {Meyer}, {Bruderer}, {Geers},
  {Pascucci}, {Lahuis}, {Juh{\'a}sz}, {Henning}, \&
  {{\'A}brah{\'a}m}}]{banzatti12a}
{Banzatti}, A., {Meyer}, M.~R., {Bruderer}, S., {et~al.} 2012, \apj, 745, 90

\bibitem[{{Baraffe} \& {Chabrier}(2010)}]{baraffe10a}
{Baraffe}, I. \& {Chabrier}, G. 2010, \aap, 521, A44

\bibitem[{{Benz} {et~al.}(2015){Benz}, {Bruderer}, {van Dishoeck}, {Melchior},
  {Wampfler}, {van der Tak}, {Goicoechea}, {Indriolo}, {Kristensen}, {Lis},
  {Mottram}, {Bergin}, {Caselli}, {Herpin}, {Hogerheijde}, {Johnstone},
  {Liseau}, {Nisini}, {Tafalla}, {Visser}, \& {Wyrowski}}]{benz15a}
{Benz}, A.~O., {Bruderer}, S., {van Dishoeck}, E.~F., {et~al.} 2015, \aap,
  subm.

\bibitem[{{Bergin} {et~al.}(2002){Bergin}, {Alves}, {Huard}, \&
  {Lada}}]{bergin02a}
{Bergin}, E.~A., {Alves}, J., {Huard}, T., \& {Lada}, C.~J. 2002, \apjl, 570,
  L101

\bibitem[{{Billot} {et~al.}(2012){Billot}, {Morales-Calder{\'o}n}, {Stauffer},
  {Megeath}, \& {Whitney}}]{billot12a}
{Billot}, N., {Morales-Calder{\'o}n}, M., {Stauffer}, J.~R., {Megeath}, S.~T.,
  \& {Whitney}, B. 2012, \apjl, 753, L35

\bibitem[{{Bisschop} {et~al.}(2006){Bisschop}, {Fraser}, {{\"O}berg}, {van
  Dishoeck}, \& {Schlemmer}}]{bisschop06a}
{Bisschop}, S.~E., {Fraser}, H.~J., {{\"O}berg}, K.~I., {van Dishoeck}, E.~F.,
  \& {Schlemmer}, S. 2006, \aap, 449, 1297

\bibitem[{{Black} \& {van Dishoeck}(1987)}]{black87a}
{Black}, J.~H. \& {van Dishoeck}, E.~F. 1987, \apj, 322, 412

\bibitem[{{Blake} {et~al.}(1995){Blake}, {Sandell}, {van Dishoeck},
  {Groesbeck}, {Mundy}, \& {Aspin}}]{blake95a}
{Blake}, G.~A., {Sandell}, G., {van Dishoeck}, E.~F., {et~al.} 1995, \apj, 441,
  689

\bibitem[{{Boogert} {et~al.}(2002){Boogert}, {Hogerheijde}, {Ceccarelli},
  {Tielens}, {van Dishoeck}, {Blake}, {Latter}, \& {Motte}}]{boogert02b}
{Boogert}, A.~C.~A., {Hogerheijde}, M.~R., {Ceccarelli}, C., {et~al.} 2002,
  \apj, 570, 708

\bibitem[{{Boogert} {et~al.}(2008){Boogert}, {Pontoppidan}, {Knez}, {Lahuis},
  {Kessler-Silacci}, {van Dishoeck}, {Blake}, {Augereau}, {Bisschop},
  {Bottinelli}, {Brooke}, {Brown}, {Crapsi}, {Evans}, {Fraser}, {Geers},
  {Huard}, {J{\o}rgensen}, {{\"O}berg}, {Allen}, {Harvey}, {Koerner}, {Mundy},
  {Padgett}, {Sargent}, \& {Stapelfeldt}}]{boogert08a}
{Boogert}, A.~C.~A., {Pontoppidan}, K.~M., {Knez}, C., {et~al.} 2008, \apj,
  678, 985

\bibitem[{{Boogert} {et~al.}(2000){Boogert}, {Tielens}, {Ceccarelli},
  {Boonman}, {van Dishoeck}, {Keane}, {Whittet}, \& {de Graauw}}]{boogert00a}
{Boogert}, A.~C.~A., {Tielens}, A.~G.~G.~M., {Ceccarelli}, C., {et~al.} 2000,
  \aap, 360, 683

\bibitem[{{Butner} {et~al.}(1995){Butner}, {Lada}, \& {Loren}}]{butner95a}
{Butner}, H.~M., {Lada}, E.~A., \& {Loren}, R.~B. 1995, \apj, 448, 207

\bibitem[{{Caselli} {et~al.}(1995){Caselli}, {Myers}, \&
  {Thaddeus}}]{caselli95a}
{Caselli}, P., {Myers}, P.~C., \& {Thaddeus}, P. 1995, \apjl, 455, L77

\bibitem[{{Connelley} \& {Greene}(2010)}]{connelley10a}
{Connelley}, M.~S. \& {Greene}, T.~P. 2010, \aj, 140, 1214

\bibitem[{{Cook} {et~al.}(2011){Cook}, {Whittet}, {Shenoy}, {Gerakines},
  {White}, \& {Chiar}}]{cook11a}
{Cook}, A.~M., {Whittet}, D.~C.~B., {Shenoy}, S.~S., {et~al.} 2011, \apj, 730,
  124

\bibitem[{{Dalgarno}(2006)}]{dalgarno06a}
{Dalgarno}, A. 2006, Proc.\ Natl.\ Acad.\ Sci.\ USA, 103, 12269

\bibitem[{{Daniel} {et~al.}(2005){Daniel}, {Dubernet}, {Meuwly}, {Cernicharo},
  \& {Pagani}}]{daniel05a}
{Daniel}, F., {Dubernet}, M.-L., {Meuwly}, M., {Cernicharo}, J., \& {Pagani},
  L. 2005, \mnras, 363, 1083

\bibitem[{{de Graauw} {et~al.}(2010){de Graauw}, {Helmich}, {Phillips},
  {Stutzki}, {Caux}, {Whyborn}, {Dieleman}, {Roelfsema}, {Aarts}, {Assendorp},
  {Bachiller}, {Baechtold}, {Barcia}, {Beintema}, {Belitsky}, {Benz}, {Bieber},
  {Boogert}, {Borys}, {Bumble}, {Ca{\"i}s}, {Caris}, {Cerulli-Irelli},
  {Chattopadhyay}, {Cherednichenko}, {Ciechanowicz}, {Coeur-Joly}, {Comito},
  {Cros}, {de Jonge}, {de Lange}, {Delforges}, {Delorme}, {den Boggende},
  {Desbat}, {Diez-Gonz{\'a}lez}, {di Giorgio}, {Dubbeldam}, {Edwards},
  {Eggens}, {Erickson}, {Evers}, {Fich}, {Finn}, {Franke}, {Gaier}, {Gal},
  {Gao}, {Gallego}, {Gauffre}, {Gill}, {Glenz}, {Golstein}, {Goulooze},
  {Gunsing}, {G{\"u}sten}, {Hartogh}, {Hatch}, {Higgins}, {Honingh}, {Huisman},
  {Jackson}, {Jacobs}, {Jacobs}, {Jarchow}, {Javadi}, {Jellema}, {Justen},
  {Karpov}, {Kasemann}, {Kawamura}, {Keizer}, {Kester}, {Klapwijk}, {Klein},
  {Kollberg}, {Kooi}, {Kooiman}, {Kopf}, {Krause}, {Krieg}, {Kramer},
  {Kruizenga}, {Kuhn}, {Laauwen}, {Lai}, {Larsson}, {Leduc}, {Leinz}, {Lin},
  {Liseau}, {Liu}, {Loose}, {L{\'o}pez-Fernandez}, {Lord}, {Luinge}, {Marston},
  {Mart{\'{\i}}n-Pintado}, {Maestrini}, {Maiwald}, {McCoey}, {Mehdi}, {Megej},
  {Melchior}, {Meinsma}, {Merkel}, {Michalska}, {Monstein}, {Moratschke},
  {Morris}, {Muller}, {Murphy}, {Naber}, {Natale}, {Nowosielski}, {Nuzzolo},
  {Olberg}, {Olbrich}, {Orfei}, {Orleanski}, {Ossenkopf}, {Peacock}, {Pearson},
  {Peron}, {Phillip-May}, {Piazzo}, {Planesas}, {Rataj}, {Ravera}, {Risacher},
  {Salez}, {Samoska}, {Saraceno}, {Schieder}, {Schlecht}, {Schl{\"o}der},
  {Schm{\"u}lling}, {Schultz}, {Schuster}, {Siebertz}, {Smit}, {Szczerba},
  {Shipman}, {Steinmetz}, {Stern}, {Stokroos}, {Teipen}, {Teyssier}, {Tils},
  {Trappe}, {van Baaren}, {van Leeuwen}, {van de Stadt}, {Visser}, {Wildeman},
  {Wafelbakker}, {Ward}, {Wesselius}, {Wild}, {Wulff}, {Wunsch}, {Tielens},
  {Zaal}, {Zirath}, {Zmuidzinas}, \& {Zwart}}]{degraauw10a}
{de Graauw}, T., {Helmich}, F.~P., {Phillips}, T.~G., {et~al.} 2010, \aap, 518,
  L6

\bibitem[{{Devine} {et~al.}(1997){Devine}, {Bally}, {Reipurth}, \&
  {Heathcote}}]{devine97a}
{Devine}, D., {Bally}, J., {Reipurth}, B., \& {Heathcote}, S. 1997, \aj, 114,
  2095

\bibitem[{{Dunham} {et~al.}(2014){Dunham}, {Stutz}, {Allen}, {Evans},
  {Fischer}, {Megeath}, {Myers}, {Offner}, {Poteet}, {Tobin}, \&
  {Vorobyov}}]{dunham14a}
{Dunham}, M.~M., {Stutz}, A.~M., {Allen}, L.~E., {et~al.} 2014, Protostars and
  Planets VI, 195

\bibitem[{{Emprechtinger} {et~al.}(2009){Emprechtinger}, {Caselli}, {Volgenau},
  {Stutzki}, \& {Wiedner}}]{emprechtinger09a}
{Emprechtinger}, M., {Caselli}, P., {Volgenau}, N.~H., {Stutzki}, J., \&
  {Wiedner}, M.~C. 2009, \aap, 493, 89

\bibitem[{{Evans} {et~al.}(2009){Evans}, {Dunham}, {J{\o}rgensen}, {Enoch},
  {Mer{\'{\i}}n}, {van Dishoeck}, {Alcal{\'a}}, {Myers}, {Stapelfeldt},
  {Huard}, {Allen}, {Harvey}, {van Kempen}, {Blake}, {Koerner}, {Mundy},
  {Padgett}, \& {Sargent}}]{evans09a}
{Evans}, N.~J., {Dunham}, M.~M., {J{\o}rgensen}, J.~K., {et~al.} 2009, \apjs,
  181, 321

\bibitem[{{Evans} {et~al.}(2005){Evans}, {Lee}, {Rawlings}, \&
  {Choi}}]{evans05a}
{Evans}, II, N.~J., {Lee}, J.-E., {Rawlings}, J.~M.~C., \& {Choi}, M. 2005,
  \apj, 626, 919

\bibitem[{{Fischer} {et~al.}(2013){Fischer}, {Megeath}, {Stutz}, {Tobin},
  {Ali}, {Stanke}, {Osorio}, {Furlan}, {HOPS Team}, \& {Orion Protostar
  Survey}}]{fischer13a}
{Fischer}, W.~J., {Megeath}, S.~T., {Stutz}, A.~M., {et~al.} 2013,
  Astronomische Nachrichten, 334, 53

\bibitem[{{Flower}(1999)}]{flower99a}
{Flower}, D.~R. 1999, \mnras, 305, 651

\bibitem[{{Fraser} {et~al.}(2001){Fraser}, {Collings}, {McCoustra}, \&
  {Williams}}]{fraser01a}
{Fraser}, H.~J., {Collings}, M.~P., {McCoustra}, M.~R.~S., \& {Williams}, D.~A.
  2001, \mnras, 327, 1165

\bibitem[{{Gerakines} {et~al.}(1995){Gerakines}, {Schutte}, {Greenberg}, \&
  {van Dishoeck}}]{gerakines95a}
{Gerakines}, P.~A., {Schutte}, W.~A., {Greenberg}, J.~M., \& {van Dishoeck},
  E.~F. 1995, \aap, 296, 810

\bibitem[{{Gregersen} {et~al.}(2000){Gregersen}, {Evans}, {Mardones}, \&
  {Myers}}]{gregersen00a}
{Gregersen}, E.~M., {Evans}, II, N.~J., {Mardones}, D., \& {Myers}, P.~C. 2000,
  \apj, 533, 440

\bibitem[{{Gregersen} {et~al.}(1997){Gregersen}, {Evans}, {Zhou}, \&
  {Choi}}]{gregersen97a}
{Gregersen}, E.~M., {Evans}, II, N.~J., {Zhou}, S., \& {Choi}, M. 1997, \apj,
  484, 256

\bibitem[{{Hasegawa} \& {Herbst}(1993)}]{hasegawa93a}
{Hasegawa}, T.~I. \& {Herbst}, E. 1993, \mnras, 261, 83

\bibitem[{{Herbig}(1977)}]{herbig77a}
{Herbig}, G.~H. 1977, \apj, 217, 693

\bibitem[{{Hogerheijde} \& {van der Tak}(2000)}]{hogerheijde00a}
{Hogerheijde}, M.~R. \& {van der Tak}, F.~F.~S. 2000, \aap, 362, 697

\bibitem[{{Hogerheijde} {et~al.}(1999){Hogerheijde}, {van Dishoeck},
  {Salverda}, \& {Blake}}]{hogerheijde99a}
{Hogerheijde}, M.~R., {van Dishoeck}, E.~F., {Salverda}, J.~M., \& {Blake},
  G.~A. 1999, \apj, 513, 350

\bibitem[{{Ivezi{\'c}} {et~al.}(1999){Ivezi{\'c}}, {Nenkova}, \&
  {Elitzur}}]{ivezic99a}
{Ivezi{\'c}}, Z., {Nenkova}, M., \& {Elitzur}, M. 1999, {User Manual for DUSTY}
  (Univ.\ of Kentucky Internal Report, http://www.pa.uky.edu/$\sim$moshe/dusty)

\bibitem[{{Johnstone} {et~al.}(2013){Johnstone}, {Hendricks}, {Herczeg}, \&
  {Bruderer}}]{johnstone13a}
{Johnstone}, D., {Hendricks}, B., {Herczeg}, G.~J., \& {Bruderer}, S. 2013,
  \apj, 765, 133

\bibitem[{{J{\o}rgensen}(2004)}]{jorgensen04d}
{J{\o}rgensen}, J.~K. 2004, \aap, 424, 589

\bibitem[{{J{\o}rgensen} {et~al.}(2002){J{\o}rgensen}, {Sch{\"o}ier}, \& {van
  Dishoeck}}]{jorgensen02a}
{J{\o}rgensen}, J.~K., {Sch{\"o}ier}, F.~L., \& {van Dishoeck}, E.~F. 2002,
  \aap, 389, 908

\bibitem[{{J{\o}rgensen} {et~al.}(2004){J{\o}rgensen}, {Sch{\"o}ier}, \& {van
  Dishoeck}}]{jorgensen04c}
{J{\o}rgensen}, J.~K., {Sch{\"o}ier}, F.~L., \& {van Dishoeck}, E.~F. 2004,
  \aap, 416, 603

\bibitem[{{J{\o}rgensen} {et~al.}(2013){J{\o}rgensen}, {Visser}, {Sakai},
  {Bergin}, {Brinch}, {Harsono}, {Lindberg}, {van Dishoeck}, {Yamamoto},
  {Bisschop}, \& {Persson}}]{jorgensen13a}
{J{\o}rgensen}, J.~K., {Visser}, R., {Sakai}, N., {et~al.} 2013, \apjl, 779,
  L22

\bibitem[{{J{\o}rgensen} {et~al.}(2015){J{\o}rgensen}, {Visser}, {Williams}, \&
  {Bergin}}]{jorgensen15a}
{J{\o}rgensen}, J.~K., {Visser}, R., {Williams}, J., \& {Bergin}, E.~A. 2015,
  \aap, subm.

\bibitem[{{Kenyon} {et~al.}(1990){Kenyon}, {Hartmann}, {Strom}, \&
  {Strom}}]{kenyon90a}
{Kenyon}, S.~J., {Hartmann}, L.~W., {Strom}, K.~M., \& {Strom}, S.~E. 1990,
  \aj, 99, 869

\bibitem[{{Kim} {et~al.}(2011){Kim}, {Evans}, {Dunham}, {Chen}, {Lee},
  {Bourke}, {Huard}, {Shirley}, \& {De Vries}}]{kim11a}
{Kim}, H.~J., {Evans}, II, N.~J., {Dunham}, M.~M., {et~al.} 2011, \apj, 729, 84

\bibitem[{{Kim} {et~al.}(2012){Kim}, {Evans}, {Dunham}, {Lee}, \&
  {Pontoppidan}}]{kim12a}
{Kim}, H.~J., {Evans}, II, N.~J., {Dunham}, M.~M., {Lee}, J.-E., \&
  {Pontoppidan}, K.~M. 2012, \apj, 758, 38

\bibitem[{{Kristensen} {et~al.}(2012){Kristensen}, {van Dishoeck}, {Bergin},
  {Visser}, {Y{\i}ld{\i}z}, {San Jose-Garcia}, {J{\o}rgensen}, {Herczeg},
  {Johnstone}, {Wampfler}, {Benz}, {Bruderer}, {Cabrit}, {Caselli}, {Doty},
  {Harsono}, {Herpin}, {Hogerheijde}, {Karska}, {van Kempen}, {Liseau},
  {Nisini}, {Tafalla}, {van der Tak}, \& {Wyrowski}}]{kristensen12a}
{Kristensen}, L.~E., {van Dishoeck}, E.~F., {Bergin}, E.~A., {et~al.} 2012,
  \aap, 542, A8

\bibitem[{{Kryukova} {et~al.}(2012){Kryukova}, {Megeath}, {Gutermuth},
  {Pipher}, {Allen}, {Allen}, {Myers}, \& {Muzerolle}}]{kryukova12a}
{Kryukova}, E., {Megeath}, S.~T., {Gutermuth}, R.~A., {et~al.} 2012, \aj, 144,
  31

\bibitem[{{Lada}(1999)}]{lada99a}
{Lada}, C.~J. 1999, in The Origin of Stars and Planetary Systems, ed.
  {C.~J.~Lada \& N.~D.~Kylafis} (Dordrecht: Kluwer), 143

\bibitem[{{Lee}(2007)}]{lee07a}
{Lee}, J.-E. 2007, J.\ Korean Astron.\ Soc., 40, 83

\bibitem[{{Liebhart} {et~al.}(2014){Liebhart}, {G{\"u}del}, {Skinner}, \&
  {Green}}]{liebhart14a}
{Liebhart}, A., {G{\"u}del}, M., {Skinner}, S.~L., \& {Green}, J. 2014, \aap,
  570, L11

\bibitem[{{Mardones} {et~al.}(1997){Mardones}, {Myers}, {Tafalla}, {Wilner},
  {Bachiller}, \& {Garay}}]{mardones97a}
{Mardones}, D., {Myers}, P.~C., {Tafalla}, M., {et~al.} 1997, \apj, 489, 719

\bibitem[{{Maret} {et~al.}(2006){Maret}, {Bergin}, \& {Lada}}]{maret06a}
{Maret}, S., {Bergin}, E.~A., \& {Lada}, C.~J. 2006, \nat, 442, 425

\bibitem[{{McElroy} {et~al.}(2013){McElroy}, {Walsh}, {Markwick}, {Cordiner},
  {Smith}, \& {Millar}}]{mcelroy13a}
{McElroy}, D., {Walsh}, C., {Markwick}, A.~J., {et~al.} 2013, \aap, 550, A36

\bibitem[{{Minissale}(2014)}]{minissale14a}
{Minissale}, M. 2014, PhD thesis, Observatoire de Paris-Meudon \&
  Universit{\'e} de Cergy-Pontoise

\bibitem[{{Morales-Calder{\'o}n} {et~al.}(2011){Morales-Calder{\'o}n},
  {Stauffer}, {Hillenbrand}, {Gutermuth}, {Song}, {Rebull}, {Plavchan},
  {Carpenter}, {Whitney}, {Covey}, {Alves de Oliveira}, {Winston},
  {McCaughrean}, {Bouvier}, {Guieu}, {Vrba}, {Holtzman}, {Marchis}, {Hora},
  {Wasserman}, {Terebey}, {Megeath}, {Guinan}, {Forbrich}, {Hu{\'e}lamo},
  {Riviere-Marichalar}, {Barrado}, {Stapelfeldt}, {Hern{\'a}ndez}, {Allen},
  {Ardila}, {Bayo}, {Favata}, {James}, {Werner}, \& {Wood}}]{morales11a}
{Morales-Calder{\'o}n}, M., {Stauffer}, J.~R., {Hillenbrand}, L.~A., {et~al.}
  2011, \apj, 733, 50

\bibitem[{{Noble} {et~al.}(2012){Noble}, {Congiu}, {Dulieu}, \&
  {Fraser}}]{noble12a}
{Noble}, J.~A., {Congiu}, E., {Dulieu}, F., \& {Fraser}, H.~J. 2012, \mnras,
  421, 768

\bibitem[{{{\"O}berg} {et~al.}(2007){{\"O}berg}, {Fuchs}, {Awad}, {Fraser},
  {Schlemmer}, {van Dishoeck}, \& {Linnartz}}]{oberg07a}
{{\"O}berg}, K.~I., {Fuchs}, G.~W., {Awad}, Z., {et~al.} 2007, \apjl, 662, L23

\bibitem[{{Onishi} {et~al.}(2002){Onishi}, {Mizuno}, {Kawamura}, {Tachihara},
  \& {Fukui}}]{onishi02a}
{Onishi}, T., {Mizuno}, A., {Kawamura}, A., {Tachihara}, K., \& {Fukui}, Y.
  2002, \apj, 575, 950

\bibitem[{{Ott}(2010)}]{ott10a}
{Ott}, S. 2010, in ASP Conf.\ Ser. 434: Astronomical Data Analysis Software and
  Systems XIX, ed. Y.~{Mizumoto}, K.-I. {Morita}, \& M.~{Ohishi} (San
  Francisco: ASP), 139

\bibitem[{{Owen} \& {Jacquet}(2015)}]{owen15a}
{Owen}, J.~E. \& {Jacquet}, E. 2015, \mnras, 446, 3285

\bibitem[{{Padoan} {et~al.}(2014){Padoan}, {Haugb{\o}lle}, \&
  {Nordlund}}]{padoan14a}
{Padoan}, P., {Haugb{\o}lle}, T., \& {Nordlund}, {\AA}. 2014, \apj, 797, 32

\bibitem[{{Pilbratt} {et~al.}(2010){Pilbratt}, {Riedinger}, {Passvogel},
  {Crone}, {Doyle}, {Gageur}, {Heras}, {Jewell}, {Metcalfe}, {Ott}, \&
  {Schmidt}}]{pilbratt10a}
{Pilbratt}, G.~L., {Riedinger}, J.~R., {Passvogel}, T., {et~al.} 2010, \aap,
  518, L1

\bibitem[{{Pontoppidan} {et~al.}(2008){Pontoppidan}, {Boogert}, {Fraser}, {van
  Dishoeck}, {Blake}, {Lahuis}, {{\"O}berg}, {Evans}, \&
  {Salyk}}]{pontoppidan08a}
{Pontoppidan}, K.~M., {Boogert}, A.~C.~A., {Fraser}, H.~J., {et~al.} 2008,
  \apj, 678, 1005

\bibitem[{{Poteet} {et~al.}(2013){Poteet}, {Pontoppidan}, {Megeath}, {Watson},
  {Isokoski}, {Bjorkman}, {Sheehan}, \& {Linnartz}}]{poteet13a}
{Poteet}, C.~A., {Pontoppidan}, K.~M., {Megeath}, S.~T., {et~al.} 2013, \apj,
  766, 117

\bibitem[{{Qi} {et~al.}(2013){Qi}, {{\"O}berg}, {Wilner}, {D'Alessio},
  {Bergin}, {Andrews}, {Blake}, {Hogerheijde}, \& {van Dishoeck}}]{qi13a}
{Qi}, C., {{\"O}berg}, K.~I., {Wilner}, D.~J., {et~al.} 2013, Science, 341, 630

\bibitem[{{Raga} {et~al.}(2002){Raga}, {Vel{\'a}zquez}, {Cant{\'o}}, \&
  {Masciadri}}]{raga02a}
{Raga}, A.~C., {Vel{\'a}zquez}, P.~F., {Cant{\'o}}, J., \& {Masciadri}, E.
  2002, \aap, 395, 647

\bibitem[{{Rebull} {et~al.}(2014){Rebull}, {Cody}, {Covey}, {G{\"u}nther},
  {Hillenbrand}, {Plavchan}, {Poppenhaeger}, {Stauffer}, {Wolk}, {Gutermuth},
  {Morales-Calder{\'o}n}, {Song}, {Barrado}, {Bayo}, {James}, {Hora}, {Vrba},
  {Alves de Oliveira}, {Bouvier}, {Carey}, {Carpenter}, {Favata}, {Flaherty},
  {Forbrich}, {Hernandez}, {McCaughrean}, {Megeath}, {Micela}, {Smith},
  {Terebey}, {Turner}, {Allen}, {Ardila}, {Bouy}, \& {Guieu}}]{rebull14a}
{Rebull}, L.~M., {Cody}, A.~M., {Covey}, K.~R., {et~al.} 2014, \aj, 148, 92

\bibitem[{{Roelfsema} {et~al.}(2012){Roelfsema}, {Helmich}, {Teyssier},
  {Ossenkopf}, {Morris}, {Olberg}, {Shipman}, {Risacher}, {Akyilmaz},
  {Assendorp}, {Avruch}, {Beintema}, {Biver}, {Boogert}, {Borys}, {Braine},
  {Caris}, {Caux}, {Cernicharo}, {Coeur-Joly}, {Comito}, {de Lange},
  {Delforge}, {Dieleman}, {Dubbeldam}, {de Graauw}, {Edwards}, {Fich},
  {Flederus}, {Gal}, {di Giorgio}, {Herpin}, {Higgins}, {Hoac}, {Huisman},
  {Jarchow}, {Jellema}, {de Jonge}, {Kester}, {Klein}, {Kooi}, {Kramer},
  {Laauwen}, {Larsson}, {Leinz}, {Lord}, {Lorenzani}, {Luinge}, {Marston},
  {Mart{\'{\i}}n-Pintado}, {McCoey}, {Melchior}, {Michalska}, {Moreno},
  {M{\"u}ller}, {Nowosielski}, {Okada}, {Orlea{\'n}ski}, {Phillips}, {Pearson},
  {Rabois}, {Ravera}, {Rector}, {Rengel}, {Sagawa}, {Salomons},
  {S{\'a}nchez-Su{\'a}rez}, {Schieder}, {Schl{\"o}der}, {Schm{\"u}lling},
  {Soldati}, {Stutzki}, {Thomas}, {Tielens}, {Vastel}, {Wildeman}, {Xie},
  {Xilouris}, {Wafelbakker}, {Whyborn}, {Zaal}, {Bell}, {Bjerkeli}, {De Beck},
  {Cavali{\'e}}, {Crockett}, {Hily-Blant}, {Kama}, {Kaminski}, {Lefl{\'o}ch},
  {Lombaert}, {de Luca}, {Makai}, {Marseille}, {Nagy}, {Pacheco}, {van der
  Wiel}, {Wang}, \& {Y{\i}ld{\i}z}}]{roelfsema12a}
{Roelfsema}, P.~R., {Helmich}, F.~P., {Teyssier}, D., {et~al.} 2012, \aap, 537,
  A17

\bibitem[{{Safron} {et~al.}(2015){Safron}, {Fischer}, {Megeath}, {Furlan},
  {Stutz}, {Stanke}, {Billot}, {Rebull}, {Tobin}, {Ali}, {Allen}, {Booker},
  {Watson}, \& {Wilson}}]{safron15a}
{Safron}, E.~J., {Fischer}, W.~J., {Megeath}, S.~T., {et~al.} 2015, \apjl, 800,
  L5

\bibitem[{{San Jos{\'e}-Garc{\'{\i}}a} {et~al.}(2015){San
  Jos{\'e}-Garc{\'{\i}}a}, {Mottram}, {van Dishoeck}, {van der Tak},
  {Kristensen}, {Herpin}, {Chavarr{\'\i}a}, {Benz}, \& {Wyrowski}}]{sanjose14a}
{San Jos{\'e}-Garc{\'{\i}}a}, I., {Mottram}, J.~C., {van Dishoeck}, E.~F.,
  {et~al.} 2015, \aap, subm.

\bibitem[{{Schmalzl} {et~al.}(2014){Schmalzl}, {Visser}, {Walsh}, {Albertsson},
  {van Dishoeck}, {Kristensen}, \& {Mottram}}]{schmalzl14a}
{Schmalzl}, M., {Visser}, R., {Walsh}, C., {et~al.} 2014, \aap, 572, A81

\bibitem[{{Sch{\"o}ier} {et~al.}(2005){Sch{\"o}ier}, {van der Tak}, {van
  Dishoeck}, \& {Black}}]{schoier05a}
{Sch{\"o}ier}, F.~L., {van der Tak}, F.~F.~S., {van Dishoeck}, E.~F., \&
  {Black}, J.~H. 2005, \aap, 432, 369

\bibitem[{{Scholz} {et~al.}(2013){Scholz}, {Froebrich}, \& {Wood}}]{scholz13a}
{Scholz}, A., {Froebrich}, D., \& {Wood}, K. 2013, \mnras, 430, 2910

\bibitem[{{Spaans} {et~al.}(1995){Spaans}, {Hogerheijde}, {Mundy}, \& {van
  Dishoeck}}]{spaans95a}
{Spaans}, M., {Hogerheijde}, M.~R., {Mundy}, L.~G., \& {van Dishoeck}, E.~F.
  1995, \apjl, 455, L167

\bibitem[{{Stamatellos} {et~al.}(2011){Stamatellos}, {Whitworth}, \&
  {Hubber}}]{stamatellos11a}
{Stamatellos}, D., {Whitworth}, A.~P., \& {Hubber}, D.~A. 2011, \apj, 730, 32

\bibitem[{{Stamatellos} {et~al.}(2012){Stamatellos}, {Whitworth}, \&
  {Hubber}}]{stamatellos12a}
{Stamatellos}, D., {Whitworth}, A.~P., \& {Hubber}, D.~A. 2012, \mnras, 427,
  1182

\bibitem[{{Stauffer} {et~al.}(2014){Stauffer}, {Cody}, {Baglin}, {Alencar},
  {Rebull}, {Hillenbrand}, {Venuti}, {Turner}, {Carpenter}, {Plavchan},
  {Findeisen}, {Carey}, {Terebey}, {Morales-Calder{\'o}n}, {Bouvier}, {Micela},
  {Flaccomio}, {Song}, {Gutermuth}, {Hartmann}, {Calvet}, {Whitney}, {Barrado},
  {Vrba}, {Covey}, {Herbst}, {Furesz}, {Aigrain}, \& {Favata}}]{stauffer14a}
{Stauffer}, J., {Cody}, A.~M., {Baglin}, A., {et~al.} 2014, \aj, 147, 83

\bibitem[{{Tielens} {et~al.}(1991){Tielens}, {Tokunaga}, {Geballe}, \&
  {Baas}}]{tielens91a}
{Tielens}, A.~G.~G.~M., {Tokunaga}, A.~T., {Geballe}, T.~R., \& {Baas}, F.
  1991, \apj, 381, 181

\bibitem[{{Tobin} {et~al.}(2011){Tobin}, {Hartmann}, {Chiang}, {Looney},
  {Bergin}, {Chandler}, {Masqu{\'e}}, {Maret}, \& {Heitsch}}]{tobin11a}
{Tobin}, J.~J., {Hartmann}, L., {Chiang}, H.-F., {et~al.} 2011, \apj, 740, 45

\bibitem[{{Tobin} {et~al.}(2012){Tobin}, {Hartmann}, {Chiang}, {Wilner},
  {Looney}, {Loinard}, {Calvet}, \& {D'Alessio}}]{tobin12a}
{Tobin}, J.~J., {Hartmann}, L., {Chiang}, H.-F., {et~al.} 2012, \nat, 492, 83

\bibitem[{{van Dishoeck} {et~al.}(2011){van Dishoeck}, {Kristensen}, {Benz},
  {Bergin}, {Caselli}, {Cernicharo}, {Herpin}, {Hogerheijde}, {Johnstone},
  {Liseau}, {Nisini}, {Shipman}, {Tafalla}, {van der Tak}, {Wyrowski},
  {Aikawa}, {Bachiller}, {Baudry}, {Benedettini}, {Bjerkeli}, {Blake},
  {Bontemps}, {Braine}, {Brinch}, {Bruderer}, {Chavarr{\'{\i}}a}, {Codella},
  {Daniel}, {de Graauw}, {Deul}, {di Giorgio}, {Dominik}, {Doty}, {Dubernet},
  {Encrenaz}, {Feuchtgruber}, {Fich}, {Frieswijk}, {Fuente}, {Giannini},
  {Goicoechea}, {Helmich}, {Herczeg}, {Jacq}, {J{\o}rgensen}, {Karska},
  {Kaufman}, {Keto}, {Larsson}, {Lefloch}, {Lis}, {Marseille}, {McCoey},
  {Melnick}, {Neufeld}, {Olberg}, {Pagani}, {Pani{\'c}}, {Parise}, {Pearson},
  {Plume}, {Risacher}, {Salter}, {Santiago-Garc{\'{\i}}a}, {Saraceno},
  {St{\"a}uber}, {van Kempen}, {Visser}, {Viti}, {Walmsley}, {Wampfler}, \&
  {Y{\i}ld{\i}z}}]{vandishoeck11a}
{van Dishoeck}, E.~F., {Kristensen}, L.~E., {Benz}, A.~O., {et~al.} 2011,
  \pasp, 123, 138

\bibitem[{{van Kempen} {et~al.}(2009){van Kempen}, {van Dishoeck},
  {Hogerheijde}, \& {G{\"u}sten}}]{vankempen09e}
{van Kempen}, T.~A., {van Dishoeck}, E.~F., {Hogerheijde}, M.~R., \&
  {G{\"u}sten}, R. 2009, \aap, 508, 259

\bibitem[{{Visser} \& {Bergin}(2012)}]{visser12b}
{Visser}, R. \& {Bergin}, E.~A. 2012, \apjl, 754, L18 (Paper I)

\bibitem[{{Visser} {et~al.}(2011){Visser}, {Doty}, \& {van
  Dishoeck}}]{visser11a}
{Visser}, R., {Doty}, S.~D., \& {van Dishoeck}, E.~F. 2011, \aap, 534, A132

\bibitem[{{Vorobyov} {et~al.}(2013){Vorobyov}, {Baraffe}, {Harries}, \&
  {Chabrier}}]{vorobyov13a}
{Vorobyov}, E.~I., {Baraffe}, I., {Harries}, T., \& {Chabrier}, G. 2013, \aap,
  557, A35

\bibitem[{{Vorobyov} \& {Basu}(2005)}]{vorobyov05b}
{Vorobyov}, E.~I. \& {Basu}, S. 2005, \apjl, 633, L137

\bibitem[{{Vorobyov} \& {Basu}(2010)}]{vorobyov10b}
{Vorobyov}, E.~I. \& {Basu}, S. 2010, \apj, 719, 1896

\bibitem[{{Watson} \& {Salpeter}(1972)}]{watson72a}
{Watson}, W.~D. \& {Salpeter}, E.~E. 1972, \apj, 174, 321

\bibitem[{{Whittet} {et~al.}(2009){Whittet}, {Cook}, {Chiar}, {Pendleton},
  {Shenoy}, \& {Gerakines}}]{whittet09a}
{Whittet}, D.~C.~B., {Cook}, A.~M., {Chiar}, J.~E., {et~al.} 2009, \apj, 695,
  94

\bibitem[{{Wilson}(1999)}]{wilson99a}
{Wilson}, T.~L. 1999, Rep.\ Prog.\ Phys., 62, 143

\bibitem[{{Yang} {et~al.}(2010){Yang}, {Stancil}, {Balakrishnan}, \&
  {Forrey}}]{yang10a}
{Yang}, B., {Stancil}, P.~C., {Balakrishnan}, N., \& {Forrey}, R.~C. 2010,
  \apj, 718, 1062

\bibitem[{{Y{\i}ld{\i}z} {et~al.}(2012){Y{\i}ld{\i}z}, {Kristensen}, {van
  Dishoeck}, {Belloche}, {van Kempen}, {Hogerheijde}, {G{\"u}sten}, \& {van der
  Marel}}]{yildiz12a}
{Y{\i}ld{\i}z}, U.~A., {Kristensen}, L.~E., {van Dishoeck}, E.~F., {et~al.}
  2012, \aap, 542, A86

\bibitem[{{Y{\i}ld{\i}z} {et~al.}(2013){Y{\i}ld{\i}z}, {Kristensen}, {van
  Dishoeck}, {San Jos{\'e}-Garc{\'{\i}}a}, {Karska}, {Harsono}, {Tafalla},
  {Fuente}, {Visser}, {J{\o}rgensen}, \& {Hogerheijde}}]{yildiz13a}
{Y{\i}ld{\i}z}, U.~A., {Kristensen}, L.~E., {van Dishoeck}, E.~F., {et~al.}
  2013, \aap, 556, A89

\bibitem[{{Zasowski} {et~al.}(2009){Zasowski}, {Kemper}, {Watson}, {Furlan},
  {Bohac}, {Hull}, \& {Green}}]{zasowski09a}
{Zasowski}, G., {Kemper}, F., {Watson}, D.~M., {et~al.} 2009, \apj, 694, 459

\bibitem[{{Zhu} {et~al.}(2009){Zhu}, {Hartmann}, \& {Gammie}}]{zhu09a}
{Zhu}, Z., {Hartmann}, L., \& {Gammie}, C. 2009, \apj, 694, 1045

\end{thebibliography}
\end{document}